\documentclass[a4paper,12pt]{article}
\usepackage[top=3cm, bottom=3cm, left=3cm, right=3cm]{geometry} 
\usepackage{color}
\usepackage{rotating}
\usepackage{graphicx}
\usepackage{amsmath}
\usepackage{amssymb}
\usepackage{float}
\usepackage{subfig}
\usepackage[countmax]{subfloat}
\usepackage{hyperref}
\def\d{{\rm d}}
\def\displayandname#1{\rlap{$\displaystyle\csname #1\endcsname$}%
                      \qquad \texttt{\char92 #1}}

\begin{document}
\title{\bf{A Simulations Study of the Muon Response of the Iron Calorimeter
Detector at the India-based Neutrino Observatory}}

\author{Animesh Chatterjee\thanks{email: animesh@hri.res.in},
Meghna K. K.\thanks{email: meghna@imsc.res.in}, Kanishka Rawat\thanks{email:
kanishka@puhep.res.in}, Tarak Thakore\thanks{email: tarak@tifr.res.in},\\
Vipin Bhatnagar$^\ddagger$, R. Gandhi$^*$, D. Indumathi$^\dagger$,
N.K. Mondal$^\S$, Nita Sinha$^\dagger$\\
{\it $^*$Harish-Chandra Research Institute, Allahabad 211 019,
India} \\
{\it $^\dagger$The Institute of Mathematical Sciences, Chennai 600
 113, India} \\
{\it $^\ddagger$Panjab University, Chandigarh 160 014, India} \\
{\it $^\S$Tata Institute of Fundamental Research, Mumbai 400 005, India }
}
\maketitle
\begin{abstract}
{The magnetised Iron CALorimeter detector (ICAL), proposed to be built
at the India-based Neutrino Observatory (INO), is designed to study
atmospheric neutrino oscillations. The ICAL detector is optimized to
measure the muon momentum, its direction and charge. A GEANT4-based
package has been developed by the INO collaboration to simulate the ICAL
geometry and propagation of particles through the detector. The simulated
muon tracks are reconstructed using the Kalman Filter algorithm. Here
we present the first study of the response of the ICAL detector to
muons using this simulations package to determine the muon momentum
and direction resolutions as well as their reconstruction and charge
identification efficiencies. For 1--20 GeV/c muons in the central region
of the detector, we obtain an average angle-dependent momentum resolution
of 9--14\%, an angular resolution of about a degree, reconstruction
efficiency of about 80\% and a correct charge identification of about
98\%.}
\end{abstract}

\newpage

\section{Introduction}

An Iron CALorimeter (ICAL) detector is proposed to be built at
the India-based Neutrino Observatory (INO) \cite{Athar:2006yb} at Theni in
Southern India. The ICAL detector consists of 151 layers of magnetised
iron plates interleaved with Resistive Plate Chambers (RPCs) as active
detector element with a total mass of about 52 ktons. It is designed primarily
to study neutrino flavor oscillations through interactions of atmospheric
neutrinos in the detector \cite{Athar:2006yb}. The main goals of the ICAL
detector are to make precision measurements of the neutrino oscillation
parameters ($\theta_{23}$ and $|\Delta m_{32}^2|$), and more importantly,
to determine the neutrino mass hierarchy, the sign of $\Delta m_{32}^{2}$,
where $\Delta m_{32}^2 \equiv m_3^2 - m_2^2$. These goals can be
achieved by observing charged-current (CC) interactions of atmospheric
muon neutrinos and anti-neutrinos ($\nu_\mu$ and $\overline{\nu}_{\mu}$)
in the detector.

An accurate determination of the muon energy and direction is crucial to achieve the neutrino
oscillation physics goals. These muons (and sometimes also associated hadrons) arise from the CC interactions of atmospheric $\nu_\mu$ and $\overline{\nu}_\mu$ in the detector; both the $\nu_\mu \rightarrow \nu_\mu$ and $\nu_e \rightarrow \nu_\mu$ oscillation channels are sources of $\nu_\mu$ ($\overline{\nu}_\mu$). Hence the muons produced in the detector are sensitive to the pattern of neutrino flavour oscillations and in turn can be used to probe the neutrino oscillation parameters of interest. In addition, the muon energy
and direction information can be combined with that of the hadrons to
directly recover the energy and direction of the neutrinos. While CC $\nu_e$
interactions produce electrons in the final state, ICAL is not
optimised to reconstruct their energy and direction with high precision
due to their rapid energy loss mechanisms in the dense iron.

As $\nu_\mu$ and $\overline{\nu}_{\mu}$ pass through the earth
matter, they undergo different interactions depending on the mass
hierarchy. Hence, the hierarchy may be determined by observing individual
event rates for $\nu_\mu$ and $\overline{\nu}_{\mu}$, which give rise
to $\mu^-$ and $\mu^+$, respectively, in CC interactions. Therefore, the
ICAL detector must be optimised to correctly identify the charge of muons.

In this paper, we present for the first time a GEANT4-based simulation study
of the response of the ICAL detector to muons. The atmospheric
neutrino flux peaks at a few hundred MeVs, and then falls rapidly
with neutrino energy with roughly an $E^{-2.7}$ dependence. In contrast,
the interaction cross-section only increases linearly with energy. Hence events in
the detector are dominantly of lower energy, below 20 GeV. We therefore study
the reconstruction efficiency, momentum resolution, angular
resolution and the charge identification efficiency of muons in the
(1--20) GeV/c momentum regime.

The paper is organised as follows: in Section 2 we describe the
ICAL detector geometry and the nature of the magnetic field applied in
the detector. In Section 3 we briefly explain our procedure to simulate
muon tracks in the detector and algorithms to reconstruct their momentum
and charge. The Kalman filter is briefly described in Appendix A. We use these algorithms in Section 4 to obtain the response of
the ICAL detector to fixed-energy, fixed-direction muons. In particular,
we present here the up/down discrimination ability and angle
resolution for a sample of 10000 muons with fixed energy and zenith
angle. We also study the muon momentum resolution in certain
azimuthal angle bins, depending on the magnetic field distribution.
(Some details on the choice of azimuthal bins are presented
in Appendix B.) We then determine the azimuthal-averaged momentum
reconstruction efficiencies and charge-identification (CID)
efficiency of muons. The detector response to muons has been already
used to study the physics reach of ICAL with respect to precision
measurements of neutrino oscillation parameters as well as determination
of the neutrino mass hierarchy \cite{physics},\cite{physics1}. We conclude
with discussions in Section 5.

\section{ICAL Detector Geometry and Magnetic Field}

\subsection{Detector Geometry}

The proposed ICAL detector has a modular structure and its geometry
has been simulated\footnote{Details of implementation of the different
parameters of the ICAL detector in the Geant4 based code of ICAL will be
presented in Ref.\cite{asmita}} using the GEANT4 package available
from the European Center for Nuclear Research (CERN) \cite{geant}. The
full detector consists of three modules, each of size 16~m $\times$
16~m $\times$ 14.45~m, placed along the $x$ direction, with a gap of
20~cm between them. The direction along which the modules are placed is
labelled as the $x$-direction with the remaining horizontal transverse
direction being labelled $y$. The $z$-axis points vertically upwards
so that the polar angle equals the zenith angle $\theta$. The origin is
taken to be the centre of the central module.  Each module comprises 151
horizontal layers of 5.6~cm thick iron plates of size 16~m $\times$ 16~m
$\times$ 5.6~cm with a vertical gap of 4~cm, interleaved with RPCs. The
basic RPC units of size 1.84~m $\times$ 1.84~m $\times$ 2.5~cm are
placed in grid-format within the air gap, with a 16~cm horizontal gap
between them accommodating steel support structures in both $x$ and $y$
directions. Hence the iron sheets are supported every 2 m in both the $x$
and $y$ directions.

Vertical slots at {{$x=x_0 \pm 400$ cm (where $x_0$ is the central $x$
value of each module)}} extending up to $y=\pm 400$ cm and
cutting through all layers are provided to accommodate the four copper
coils that wind around the iron plates, providing an $x$-$y$ magnetic
field as shown in Fig.~\ref{fig:magfield}. In the central region
{{of each
module}}, typical values of the field strength are about 1.5 T in
the $y$-direction as obtained from simulations using MAGNET6.26 software
\cite{magnetcode}. The detector excluding the coils weighs about 52
kton, with 98\% iron where neutrino interactions are dominantly expected
to occur, and less than 2\% glass of the RPCs.

\subsection{The Magnetic Field}

\begin{figure}[htp]
\renewcommand{\figurename}{Fig.}
\begin{center}\includegraphics[width=0.65\textwidth]{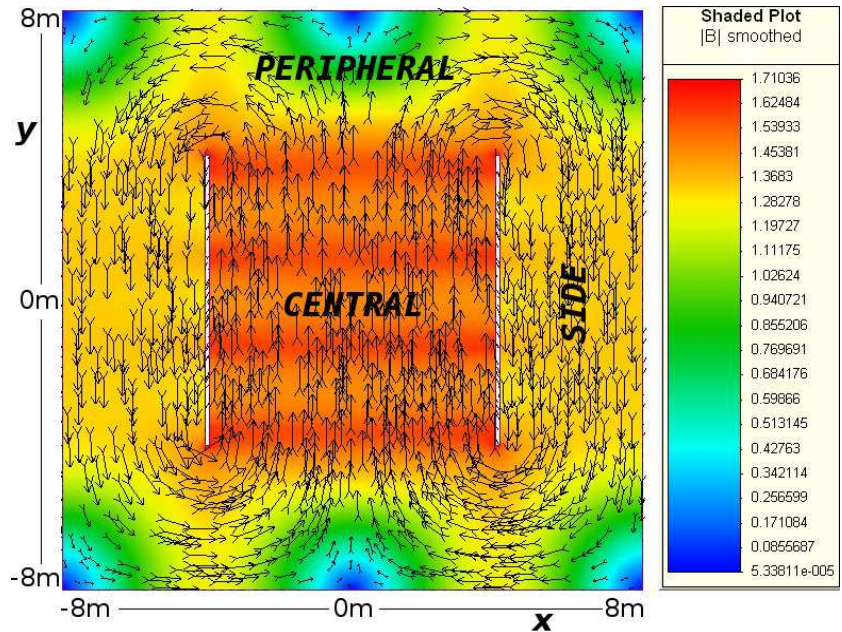}
\end{center}
\caption{Magnetic field map in the central plate of the central module
($z=0$), as generated by the MAGNET6 software. The length and direction
of the arrows indicate the magnitude and direction of the field; the
magnitude is also shown (in T) according to the colour-coding indicated
on the right.}
\label{fig:magfield}
\end{figure}

The magnetic field lines in a single iron plate {{in the
central module}}
are shown in Fig.~\ref{fig:magfield}. The arrows denote the direction
of magnetic field lines while the length of the arrows (or the
shading) indicates the magnitude of the field. Notice that the field
direction reverses on the two sides of the coil slots (beyond
{{$x_0\pm$4~m)}}
in the $x$-direction. In between the coil slots ({{an 8~m
$\times$ 8~m}} square
area in the $x$-$y$ plane) the field is maximum and nearly uniform in
both magnitude {{(to about 10\%)}} and direction; we refer to this as the {\it central region}. Near the
edges
in the $x$ direction (outside the coil slots) the field is
{{also fairly}}
uniform, but in the opposite direction; this is called the {\it side
region}. Near the edges in the $y$ direction, i.e., in the regions
{{$4$~m $\le \vert y \vert \le 8$~m}}, both the direction
and magnitude of the magnetic field vary considerably; this region is
labelled as the {\it peripheral region}.

{{The field has been generated
in the centre of the iron plate, viz., at $z=0$, and has been assumed
uniform over the entire thickness of the iron plate at every $(x,y)$
position. Since the field falls off by many orders of magnitude in
the 4~cm air gap between the iron plates, it is taken to be zero in
these regions. The magnetic field is also taken to be zero in the (non-magnetic)
steel support structures. Therefore there is neither a magnetic field
nor an active detector element in the region of the support structures;
these, along with the coil slot, form the bulk of the dead spaces of
the detector.}}

The side and peripheral regions are beset by edge effects as well as
by rapidly changing and smaller magnitude of the magnetic field; these
regions will be studied separately in order to understand the impact
on fiducial volume, etc. We confine ourselves, in the present study,
to tracks generated only in the central region ($-4$ m $\le x \le 4$ m
and $-4$ m $\le y \le 4$ m), although the particle may subsequently travel
outside this region or even exit the detector. The magnetic field thus
breaks the azimuthal symmetry of the detector geometry.

\subsection{The Active Detector Elements, RPCs}

In order to appreciate the hit pattern in the simulated detector
it is necessary to describe the active detector elements, the RPCs.
These are glass chambers made by sealing two 3 mm thick glass sheets
with a uniform gap of 2 mm using plastic edges and spacers through which
a mixture of R134A ($\sim$ 95\%), isobutane and trace amounts of SF$_6$
gas continually flows{{, with a high DC voltage across them}}. In brief, the
working principle of an RPC is the ionisation of the gaseous medium when
a charged particle passes through it. The combination of gases keeps the
signal localised and the location is used to track the charged particle
through the detector. For more details, see Ref.~\cite{rpc_char}.

The component most relevant to the simulation and track reconstruction
is a 150 micron thick copper sheet pasted on {{the {\it
inside} of a 5 mm
thick foam (for structural strength and electrical insulation) placed
both above and below the glass chamber. The copper sheet (pasted on the
side of the foam facing the glass) is insulated from the
glass by a few sheets of mylar and is used to inductively pick up the
signal when a charged particle traverses the chamber.}} This layer is
scored through with grooves to form strips of width 1.96 cm in such a
way that the strips above and below are transverse to each other, that
is, in the $x$ and $y$ directions. These pick-up strips thus provide the
$x$ and $y$ location of the charged particle as it traverses the RPC. A
timing resolution of about 1.0 ns is assumed as also an efficiency of
95\%, consistent with observations of RPCs that have been built as part
of the R \& D for ICAL \cite{rpc_char}.

\section{ICAL Muon Simulation Framework}

The ICAL detector is most sensitive to
muons. Muons being minimum ionizing particles leave long, clean,
tracks in the detector. The muon momentum can be determined from the
curvature of its track as it propagates in the magnetized detector and
also by measuring its path length. Due to the precise (nanosecond) response
time of RPCs, the up-coming muons can be distinguished from down-going
muons. This is essential since neutrinos produced in the atmosphere on
the other side of the Earth have different oscillation signatures at
the detector than those produced on the same side. The charge of the
muon can also be found from the direction of curvature of the track
and this in turn distinguishes $\nu_\mu$ and $\overline{\nu}_\mu$
interactions. In contrast, the energy of hadrons is determined from the
total number of hits in a hadron shower \cite{hadronresponse}.

\begin{figure}[htp]
\renewcommand{\figurename}{Fig.}
\begin{center}
\includegraphics[width=0.7\textwidth]{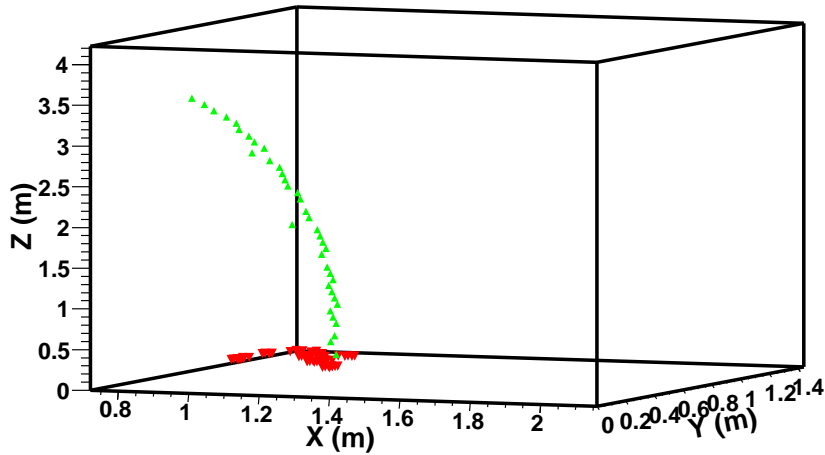}
\end{center}
\caption{{Sample track of a 10 GeV neutrino event generating a
5.1 GeV muon track and associated hadron shower in the ICAL detector. The
$X$, $Y$, and $Z$ axes are in m and the region containing the track
corresponds to the central region of the detector.}}
\label{fig:samplevicetrack}
\end{figure}

A typical neutrino CC interaction giving rise to an event
with a muon track and associated hadron shower is shown in
Fig.~\ref{fig:samplevicetrack}. It can be seen that the muon track is
clean with typically 1 hit per layer. In the case of multiple tracks, the
reconstructed track closest to the vertex is considered as the muon track.

The muon track reconstruction is based on a Kalman filter algorithm
\cite{kalman} that takes into account the local magnetic field. The
structure of each of these algorithms is briefly described below. Details
of the reconstruction code can be found elsewhere \cite{asmita}.

\subsection{Hit and Track Generation}

When a charged particle, for example, a muon, goes through an RPC, it
gives a signal which is counted as a ``hit'' in the detector with assigned $x$ or $y$ values from the respective pick-up strip
information, a $z$-value from the layer information, and a time stamp
$t$. 

The hit information is digitised, keeping in mind the strip
width so that the spatial resolution in the horizontal plane is of the
order of cm (due to the strip width) while that in the $z$ direction is
of the order of mm (due to the gas gap between the glass plates in the
RPCs). On-going studies of RPCs \cite{rpc_char} in
the INO labs have also given information on cross-talk, the probability
of both adjacent strips giving signals in the detector. This has also
been incorporated into the analysis. In addition, since the $x$ and $y$
strip information are independent, all the possible pairs of nearby $x$ and $y$ hits in a plane
are combined to form a cluster. A set of clusters
generated in a few successive layers is called a tracklet.

\paragraph{The track finder}: Clusters form the basic elements for the
track finder algorithm. Typically, muons leave only about one hit per
layer they traverse ($\sim 1.6$, on average) while hadrons and electrons,
due to the very different nature of their energy loss mechanisms, give
showers which leave several hits per layer. Including the possibility
of cross-talk (which gives the possibility of more than one hit in a
plane for muons as well), the separation/rejection of hadronic showers
is done by an algorithm that uses a cut on the total number of hits in
a given RPC module. 

The track finder algorithm uses a simple curve fitting algorithm to find
possible tracklets by finding clusters in three adjacent planes.
The finder includes the possibility of no hit (due to inefficiency) in a given plane; in this case it considers the adjoining planes as well.
Adjacent tracklets are associated into tracks and the longest possible
tracks are found \cite{Marshall} by iteration of this process.
The direction (up/down) of the track is
calculated from the timing information which is averaged over the $x$
and $y$ timing values in a plane. The track finder also separates
out tracks as ``shower-like" or ``muon-like"; for the case of muon-like
tracks {{which have at least 5 hits in the event}}, the clusters in a layer
are averaged to yield a single hit per layer with $x$, $y$ and timing
information and are sent to the track fitter for further analysis.
{{This
translates to a minimum momentum of about 0.4 GeV/c for a nearly vertical
muon ($0.4/\cos\theta$ in the absence of the magnetic field) below which
no track is fitted.}}

\paragraph{The track fitter}: A Kalman-filter based algorithm is used to
fit the tracks based on the bending of the tracks in the magnetic
field. Every track is identified by a starting vector $X_0 = (x, y,
\d x/\d z, \d y/\d z, q/p)$ which contains the position of the earliest
hit $(x,y,z)$ as recorded by the finder, with the charge-weighted inverse
momentum $q/p$ taken to be zero. Since the tracks are virtually straight
in the starting section, the initial track direction is calculated from
the first two layers. This initial state vector is then extrapolated to
the next layer using a standard Kalman-filter based algorithm described
briefly in Appendix A, using the information on the local magnetic field
and the geometry and composition of the matter through which the particle
propagates. The state extrapolation takes into account process noise due
to multiple scattering as described in Ref.~\cite{Wolin} and energy loss
in matter, mostly iron, according to the Bethe formula
\cite{Bethe}. The extrapolated point is compared with the actual location
of a hit in that layer, if any, and the process is iterated.

As stated earlier, clusters are first associated into tracklets. The
ends of the overlapping tracklets are matched to form a single longer track.
Typically, tracks from charged current muon neutrino interactions in ICAL
have a single long track due to the muons and a shower from the hadrons
near the vertex. There are rarely two or more tracks. The track closest
to the vertex is then identified as the muon track. For single muons in
the detector, as in the studies presented here, the shower is absent.

\paragraph{The fit parameters}: The process of iteration also achieves
the best fit to the track. The track is then extrapolated to another
half-layer of iron (since the interaction is most likely to have taken
place in the iron) to determine the vertex of the interaction and the best
fit value of the momentum at the vertex is returned as the reconstructed
momentum (both in magnitude and direction). While $q/p$ determines the
magnitude of the momentum at the vertex, the direction is reconstructed
using $\d x/\d z$ and $\d y/\d z$, which yield $\cos\theta$ and $\phi$.

Only fits for which the quality of fit is better than $\chi^2/\hbox{ndf}
<10$ are used in the analysis (similar results are obtained with a
tighter selection criterion).

Each muon track is further analysed in order to identify
its direction, charge and momentum. We present the response of
the ICAL detector to muons in the next section.

\section{Response of ICAL to Muons}

We propagate 10000 muons uniformly from a vertex randomly located in
the $ 8~m \times 8~m \times 10~m $ volume {{which comprises
the central region of the central module of the detector}} where the
magnetic field is uniform. For this analysis we have considered events
with the $z$ coordinate of the input vertex position to be $z_{\rm in}
\le 400$ cm. Only the vertex position is confined to the central region;
the particle, by virtue of its momentum and position, may travel outside
this region and exit the detector altogether as well. The effect of
the peripheral region of the detector, including the edge events, will
be presented elsewhere \cite{peripheral}. {{While the input
momentum and zenith angle are kept fixed in each case, the azimuthal angle
is uniformly average over the entire range $-\pi \le \phi \le \pi$.}}

In each case, we study the number of tracks reconstructed, the direction
reconstruction, including up/down discrimination, and zenith angle
resolution. Next we go on to study the muon momentum resolution; an
additional selection criterion is required in order to obtain good
momentum resolution. Furthermore, we will find that the azimuthal
dependence is non-trivial. The momentum reconstruction efficiency and
the relative charge identification efficiency are computed for those
events that pass the additional selection criterion and are discussed
at the end of the section.

\subsection{{Track Reconstruction Efficiency}}

\begin{figure}[htp]
\renewcommand{\figurename}{Fig.}
  \centering
\includegraphics[width=0.5\textwidth]{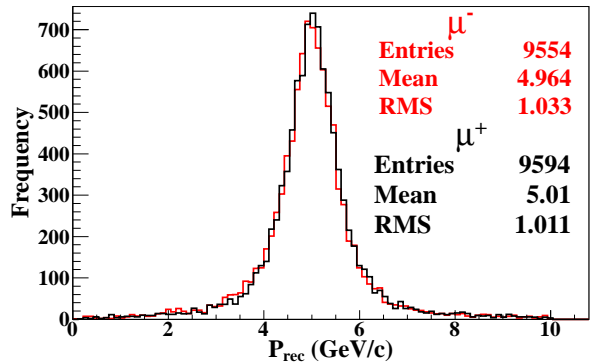}
\caption{Reconstructed momentum distributions for $(P_{\rm in},
\cos\theta) = (5\hbox{ GeV/c}, 0.65)$ smeared over the central volume
of the detector for $\mu^-$ and $\mu^+$ particles.}
\label{fig:mu+-}
\end{figure}

Separate analyses were done for $\mu^{-}$ and $\mu^{+}$. In
Fig.~\ref{fig:mu+-} the histogram of reconstructed momentum $P_{\rm rec}$
GeV/c is plotted for tracks with quality of fit such that
$\chi^2/{\hbox{ndf}} <10$, for input values $(P_{\rm in},\cos\theta) = (5\hbox{ GeV/c}, 0.65)$.
The ratio of successfully reconstructed tracks to the total number of
events is the track reconstruction efficiency. The results for $\mu^-$
and $\mu^+$ are virtually identical as seen from Fig.~\ref{fig:mu+-}.
Hence, only results for $\mu^{-}$ are presented in what follows.

{{Typically (except for very horizontal events with $\cos\theta < 0.4$),
the track reconstruction efficiency for events with momentum $P_{\rm in}
> 2$ GeV/c is around 90\% and rises to 95\% by 5 GeV/c or
more.}}

In addition, both fully contained and partially contained events (where
only a part of the track of a muon is contained in the detector since
it exits the detector without stopping) are considered for analysis. At
low energies, the tracks are fully contained while particles start to
leave the detector region by $P_{\rm in} \sim 6$ GeV/c, depending on
the location of the vertex and the zenith angle of the event.
{{We now
go on to study the quality of reconstruction. We begin with the up-down
discrimination ability.}} 

\subsection{Direction (Up/Down) Reconstruction}

\begin{figure}[btp]
\renewcommand{\figurename}{Fig.}
  \centering
  \subfloat[$\cos\theta_{\rm in} = 0.25$; $\theta_{\rm in}=1.318$]{\label{fig:13a}\includegraphics[width=0.48\textwidth]{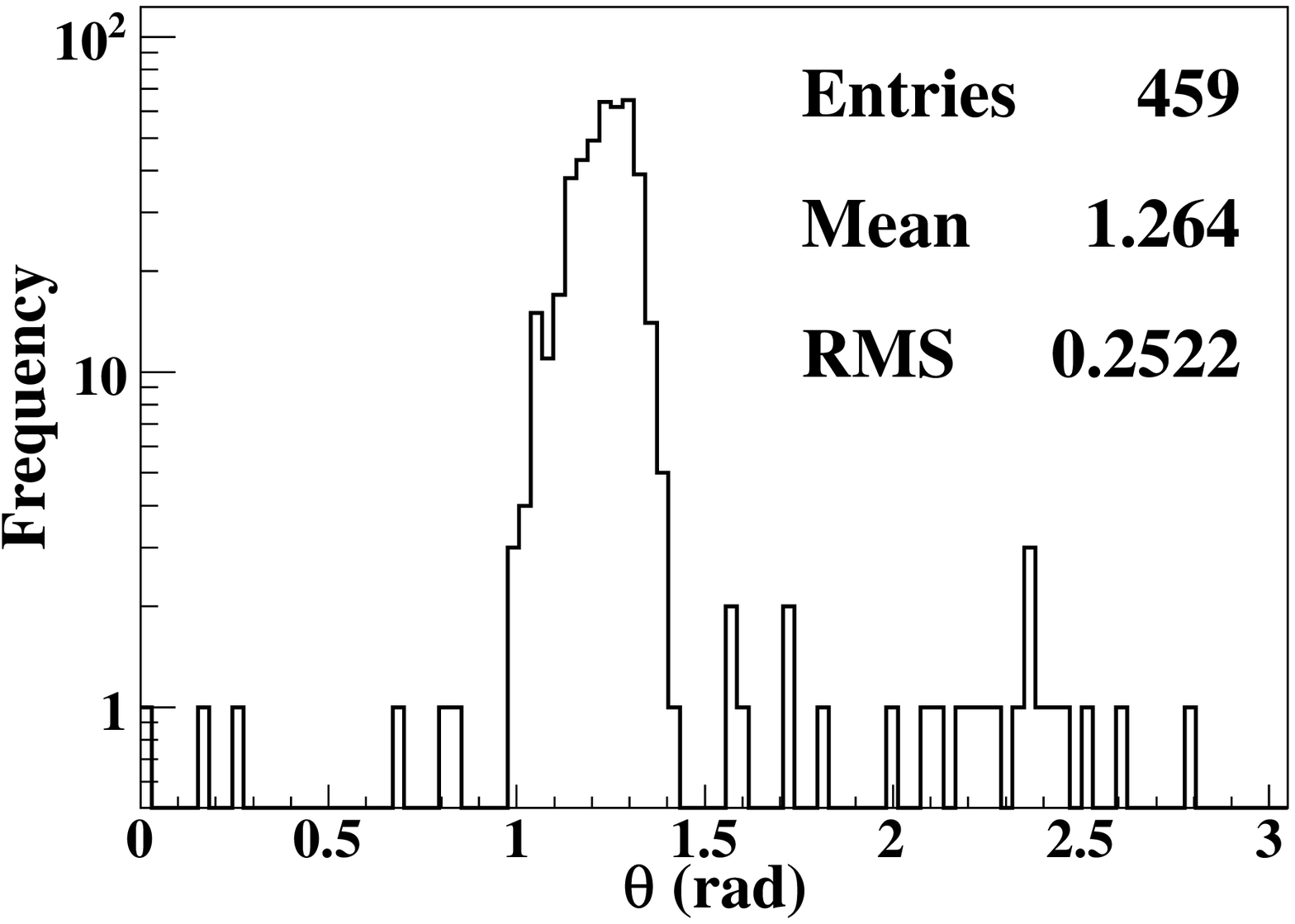}}
  \subfloat[$\cos\theta_{\rm in} = 0.85$; $\theta_{\rm in} = 0.555$]{\label{fig:13b}\includegraphics[width=0.48\textwidth]{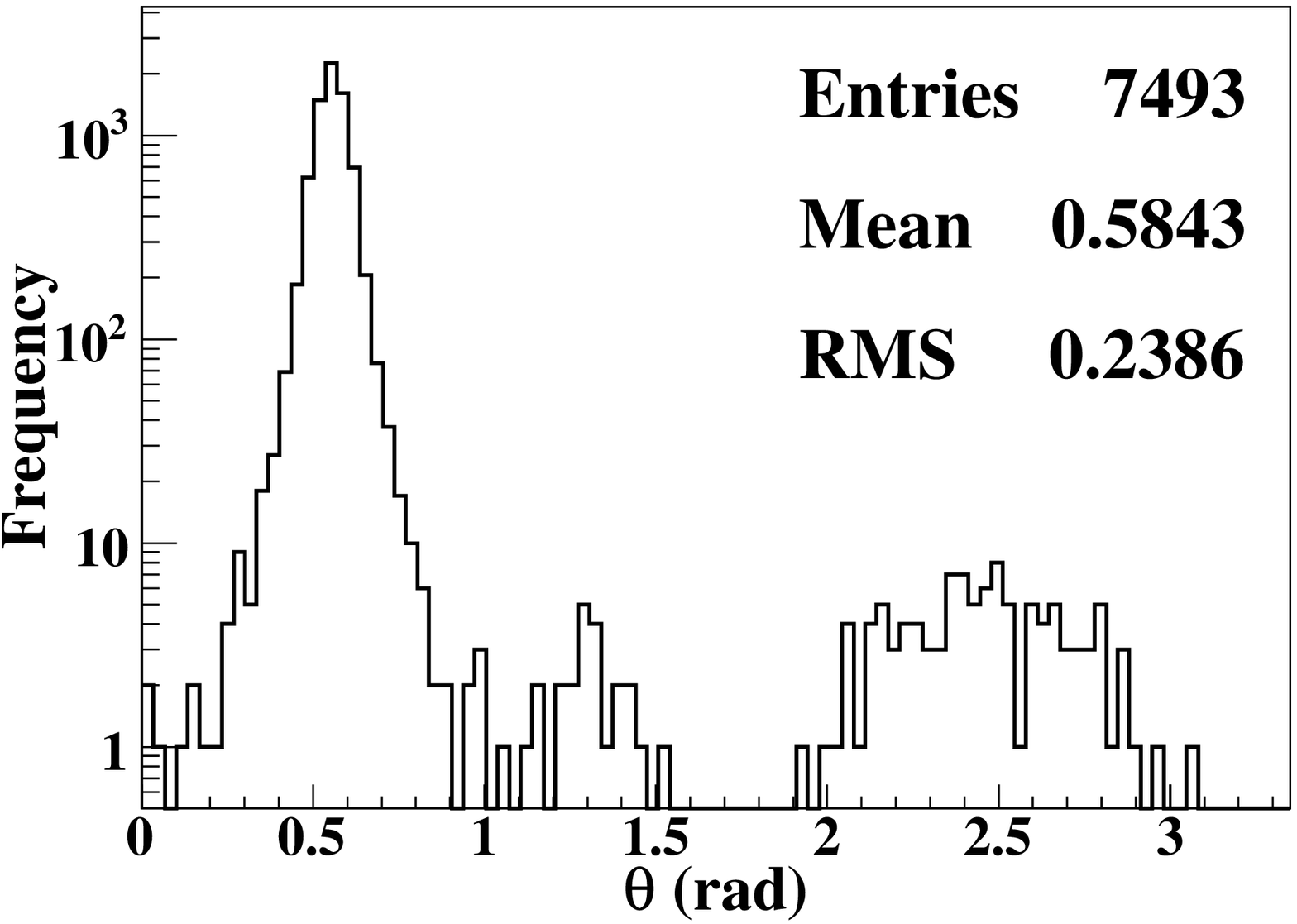}}
  \caption{Reconstructed angular distribution for $P_{\rm in} = 1$ GeV/c at two
different input angles.}
\label{fig:theta}
\end{figure}

The zenith angle is also the polar angle, $\theta$, where $\cos\theta=1
(-1)$
indicates an up-going (down-going) muon. The reconstructed zenith angle distribution
for muons with $P_{\rm in} = 1$ GeV/c at large and small angles, $\cos\theta
= 0.25$ and $\cos\theta = 0.85$, are shown in Fig.~\ref{fig:theta}.

It can be seen that a few events are reconstructed in the opposite
(downward) direction to the input direction $\theta_{\rm in}$, i.e., with
reconstructed zenith angle $\theta \sim \pi - \theta_{\rm in} $. For muons
with $P_{\rm in} = 1$ GeV/c at large (small) angles with $\cos\theta_{\rm
in} = 0.25~(0.85)$, this fraction is about 4.3 (1.5)\%. The fraction
of events reconstructed in the wrong direction drastically reduces with
energy; for example, this fraction is 0.3\% for muons with $P_{\rm in} = 2$
GeV/c at $\cos\theta_{\rm in} = 0.25$. 

{{It is seen that while the number of tracks reconstructed is also
small at large angles, it rapidly improves at smaller angles, even for
low momenta. We therefore restrict our analysis in what follows to
$\cos\theta = 0.35$ or larger.}}

\subsection{Zenith Angle Resolution}
\begin{figure}[htp]
\renewcommand{\figurename}{Fig.}
\begin{center}
\includegraphics[width=0.7\textwidth]{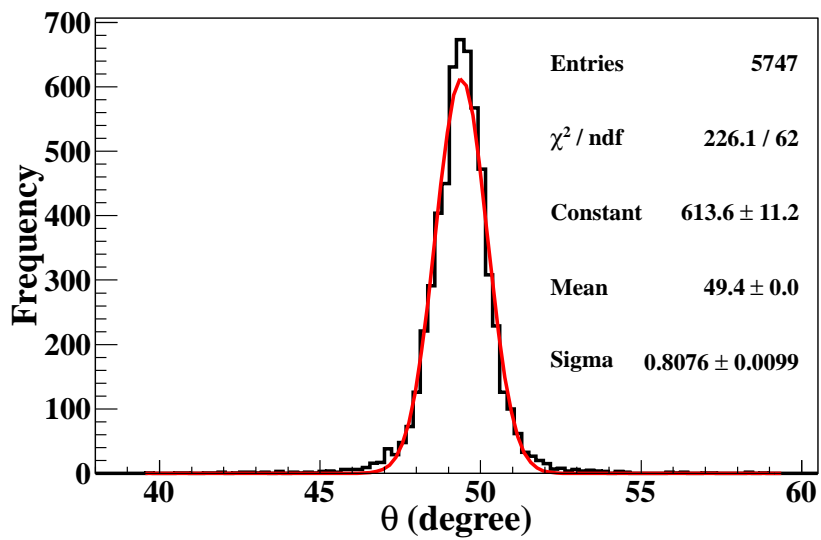}
\end{center}
\caption{Reconstructed angular distribution for input $(P_{\rm in},
\cos\theta_{\rm in}) = (5\hbox{ GeV/c}, 0.65)$; $\theta = 49.46^\circ$.}
\label{fig:5}
\end{figure}

The $\theta$ resolution is the width obtained by fitting the reconstructed
zenith angle distribution with Gaussian probability distribution
functions (pdf). The events distribution as a function of the reconstructed
$\theta$ is shown in Fig.~\ref{fig:5} for a sample input $(P_{\rm in},
\cos\theta_{\rm in}) = (5\hbox{ GeV/c}, 0.65)$. It is seen that the
distribution is very narrow, less than a degree, indicating a good
angular resolution for muons.

\begin{figure}[htp]
\renewcommand{\figurename}{Fig.}
\begin{center}
\includegraphics[width=0.7\textwidth]{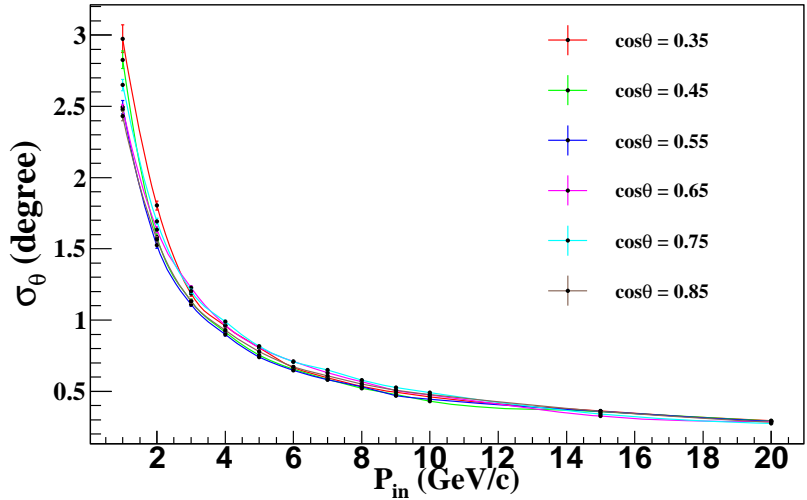}
\end{center}
\caption{Angular resolution in degrees as a function of input momentum.}
\label{fig:6}
\end{figure}

The $\theta$ resolution is shown as a function of input momentum for
different zenith angles in Fig.~\ref{fig:6}. The resolution quickly
improves with input momentum, being better than a degree for all input
angles for $P_{\rm in} > 4$ GeV/c. Beyond 10 GeV/c,
the curves approximately coincide.

{{We will see in the next section that an additional selection criterion on
the reconstructed events is required in order to obtain good momentum
resolution. However, the angle resolution remains practically the same
even when we analyse only the fraction of events whose momenta are
successfully reconstructed for their angular resolution.}}

\subsection{Muon Momentum Resolution}

The momentum resolution, $R$, is defined (with its error $\delta R$) in terms
of the RMS width $\sigma$ of the histogram of reconstructed momentum,
$P_{\rm rec}$, as,
\begin{eqnarray}
R & \equiv & \sigma/P_{\rm in}, \\ \nonumber
\delta R/R & = & \delta\sigma/\sigma~.
\end{eqnarray}

\begin{figure}[htp]
\renewcommand{\figurename}{Fig.}
  \centering
\includegraphics[width=0.8\textwidth]{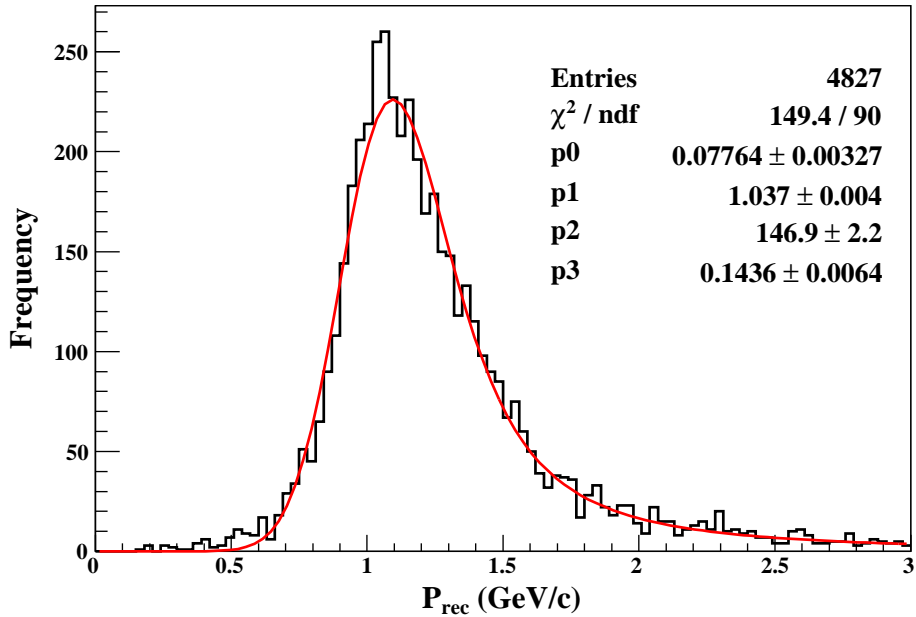}
\caption{Momentum distributions for $(P_{\rm in}, \cos\theta) =
(1\hbox{ GeV/c}, 0.65)$ fitted with Landau convoluted Gaussian; this fits
better than a pure Gaussian at lower energies, $P_{\rm in} \le 2$ GeV/c.}
\label{fig:svFit}
\end{figure}

Low energy distributions show a clear asymmetric tail (as shown in
Fig.~\ref{fig:svFit}) and so are fitted with a convolution of Landau
and Gaussian probability distribution functions (pdf). {{In the fit shown,
$p_0$ is the width (scale) parameter of the Landau density function, 
$p_1$ is the Most Probable (location) parameter of the Landau
density function, 
$p_2$ is the total area (normalization constant), and
$p_3$ is the width (sigma) of the convoluted Gaussian function.}}

For muons with
$P_{\rm in} > 2$ GeV/c, fits are with purely Gaussian pdf. In the case of
Landau-Gauss fits, the width is defined as $\sigma \equiv
\hbox{FWHM}/2.35$, where $\hbox{FWHM}=$ Full
Width at Half Maximum, in order to make a consistent and
meaningful comparison with the Gaussian fits at higher energies, where
the square root of the variance or the RMS width equals FWHM/2.35.

{{
While
the histograms at higher energies show no asymmetric tails, they fit
poorly to Gaussian distributions, with $\chi^2/{\hbox{ndf}} \ge 6$ or more.
This is because of the presence of
detector dead spaces. These dead spaces generically worsen the quality of
reconstruction and have two specific, significant effects:}}
\begin{itemize}
\item Sometimes the Kalman filter reconstruction algorithm reconstructs
two portions of a track traversing a support structure as two
disjoint tracks, with the separate track portions giving a somewhat or
significantly lower fit to the momentum depending on whether it was the
earlier or latter part of the track (since the false tracks which are only
sections of the single correct one, have arbitrary intermediate vertices
where the momentum is reconstructed). This issue is particularly severe
for small angle tracks which are more vertically oriented when the two
portions of the track on either side of the support structure may be
displaced by several layers along the $z$ axis with respect to each
other. The problem is also more severe in the case of higher energy
muons, which are more likely to traverse such spaces in the detector due
to the longer length of the tracks. For genuine neutrino events at such
energies, the hadron shower associated with these muon tracks will help
select the true vertex and reject the spurious, displaced ones. Hence
such events may be safely removed from the analysis of pure muon events;
this, while improving the momentum resolution, will worsen the momentum
reconstruction efficiency, especially at small angles.

{{This is the additional selection criterion mentioned
above.}}

\item 
Notice that the response of the detector has been averaged over the
entire azimuth. However, for the same zenith angle, $\cos\theta$, muons
with different azimuthal angles have different response. There are
several reasons for the $\phi$ dependence: the coil gaps that are
located at $x=x_0 \pm 400$ cm where $x_0$ is the centre of each module
in the $x$ direction; the support structures, which have different
dimensions in the $x$- and $y$-directions; and also the orientation of
the magnetic field. The cumulative $\phi$ dependence
is a complex consequence of all these dependences.

For instance, a muon initially directed along the $y$-axis experiences
less bending since the momentum component in the plane of the iron plates
(henceforth referred to as in-plane momentum) is parallel to the magnetic
field. Furthermore, upward-going muons that are in the negative (positive)
$x$ direction experience a force in the positive (negative) $z$ direction
(the opposite is be true for $\mu^+$) and so muons injected with $\vert
\phi \vert > \pi/2$ traverse more layers than those with the same energy
and zenith angle but with $\vert \phi \vert < \pi/2$ and hence are better
reconstructed. This is illustrated in the schematic in Fig.~\ref{fig:bend}
which shows two muons ($\mu^-$) injected at the origin with the same
momentum magnitude and zenith angle, 
one with positive momentum component in the $x$ direction,
$P_x > 0$ and the other with negative $x$ momentum component. The muon
with $P_x > 0$ (initially directed in the positive $x$ direction) bends
differently than the one with $P_x < 0$ (along negative $x$ direction)
and hence they traverse different number of layers, while having roughly
the same path length. Hence, muons with different $\phi$ elicit different
detector response.

\begin{figure}[htp]
\renewcommand{\figurename}{Fig.}
  \centering
\includegraphics[width=0.8\textwidth]{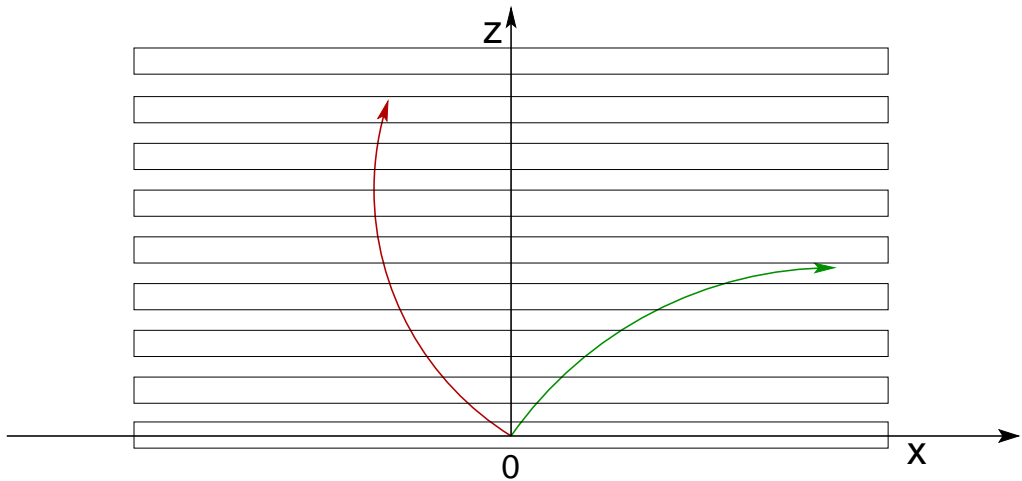}
\caption{Schematic showing muon tracks (for $\mu^-$) in the $x$-$z$ plane
for the same values of $(P_{\rm in}, \cos\theta)$ but with $\vert \phi
\vert < \pi/2$ and $> \pi/2$ (momentum component in the $x$ direction
positive and negative respectively). The different bending causes the
muon to traverse different number of layers in the two cases.}
\label{fig:bend}
\end{figure}

\end{itemize}

{{
Because of these effects, the
momentum resolution is best studied in different azimuthal angle bins
(see Appendix B for more details about the choice of the bins).}}

\subsubsection{Muon Resolution in Different Azimuthal Bins}

In all cases, we separate our muon sample into four bins: Bin I with $\vert
\phi \vert \le \pi/4$, Bin II with $\pi/4 < \vert \phi \vert \le \pi/2$,
Bin III with $\pi/2 < \vert \phi \vert \le 3\pi/4$, and Bin IV with
$3\pi/4 < \vert \phi \vert \le \pi$.

\begin{figure}[btp]
\renewcommand{\figurename}{Fig.}
  \centering
  \subfloat[$ \mid \phi \mid \le \pi/4
$]{\label{fig:55a}\includegraphics[totalheight=.18\textheight,
width=0.50\textwidth]{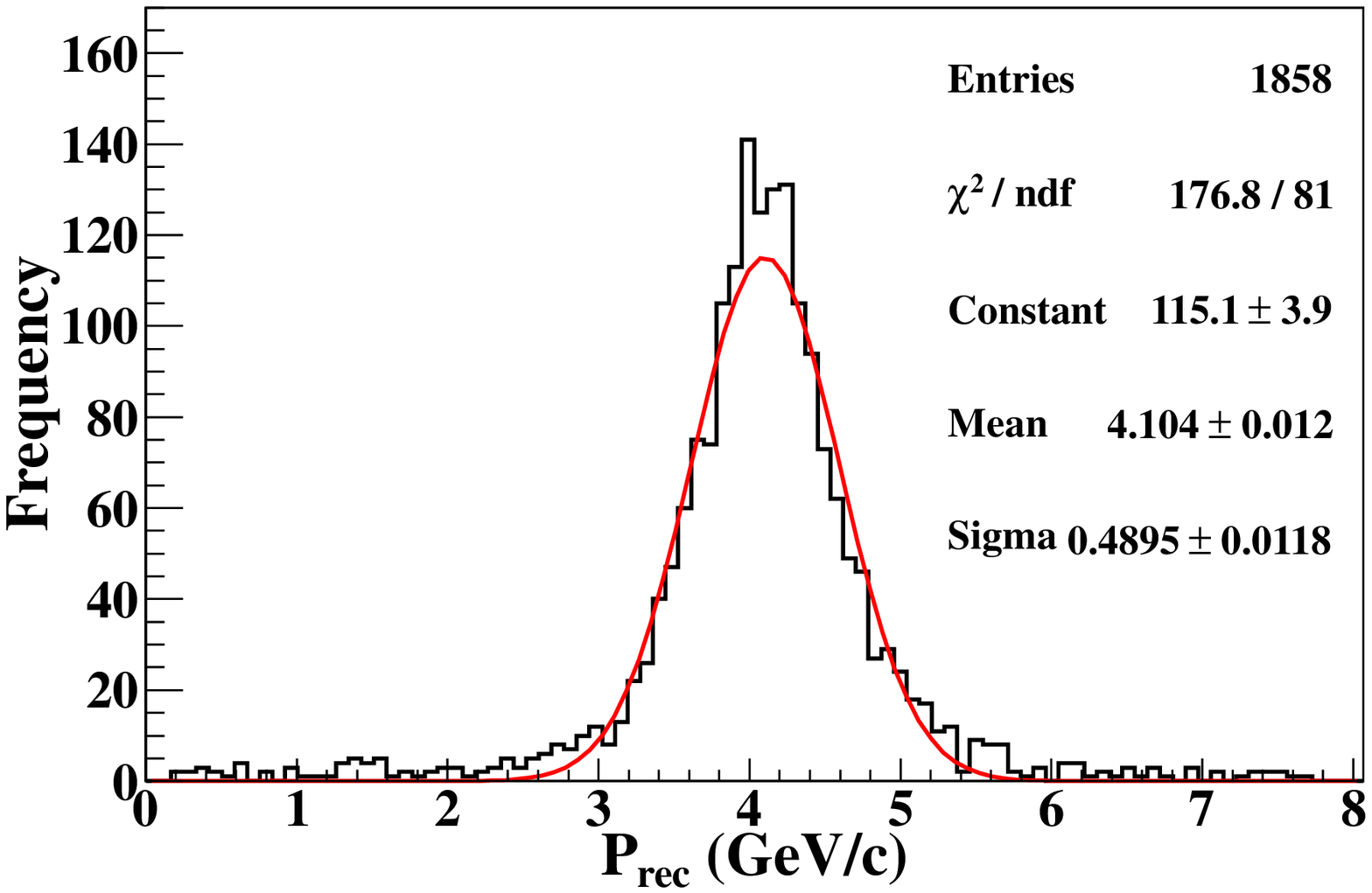}}
  \subfloat[$ \pi/4 < \mid \phi \mid \le \pi/2
$]{\label{fig:55b}\includegraphics[totalheight=.18\textheight,
width=0.50\textwidth]{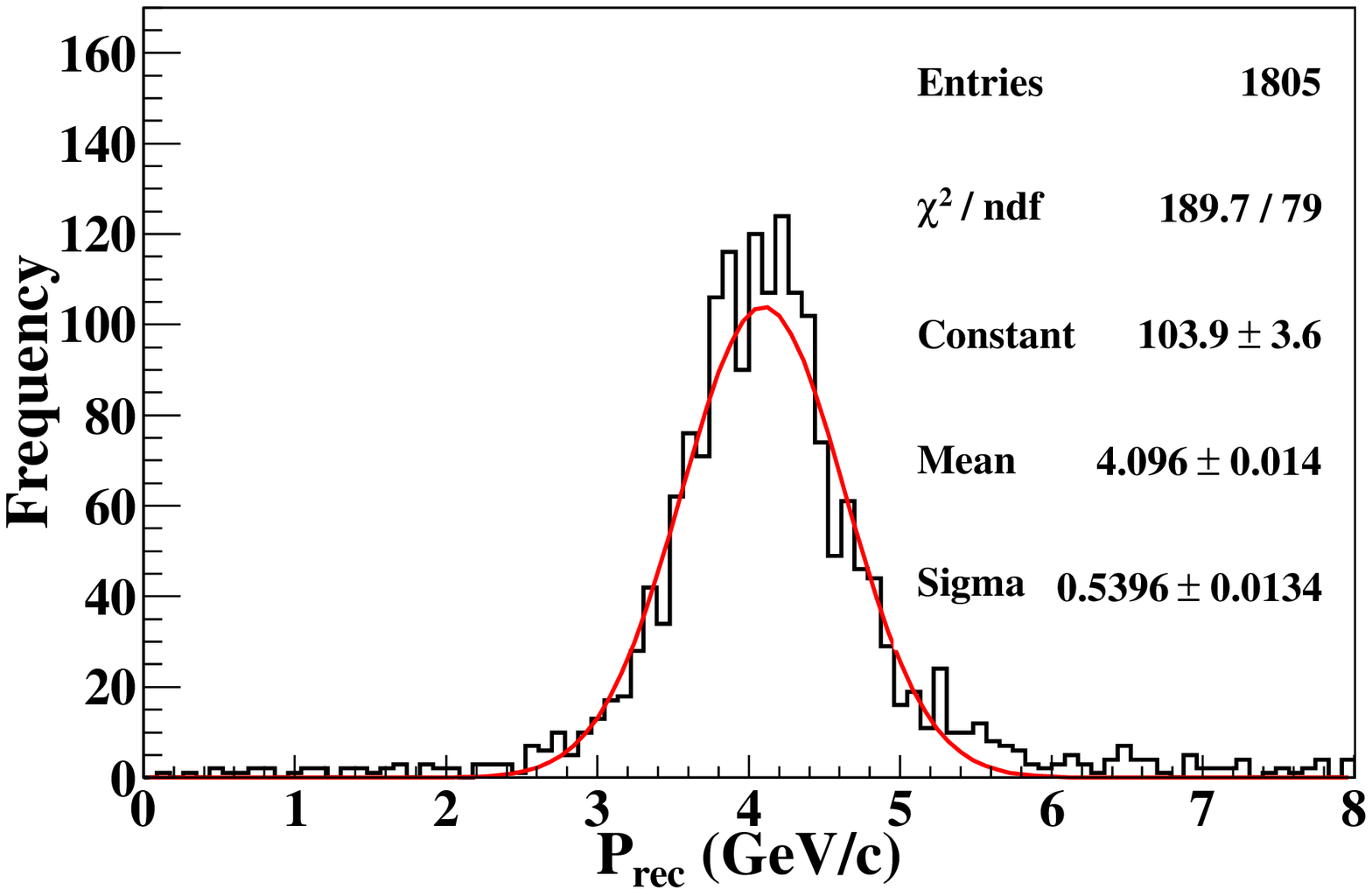}}\\
  \subfloat[$ \pi/2 < \mid \phi \mid \le 3\pi/4
$]{\label{fig:55c}\includegraphics[totalheight=.18\textheight,
width=0.50\textwidth]{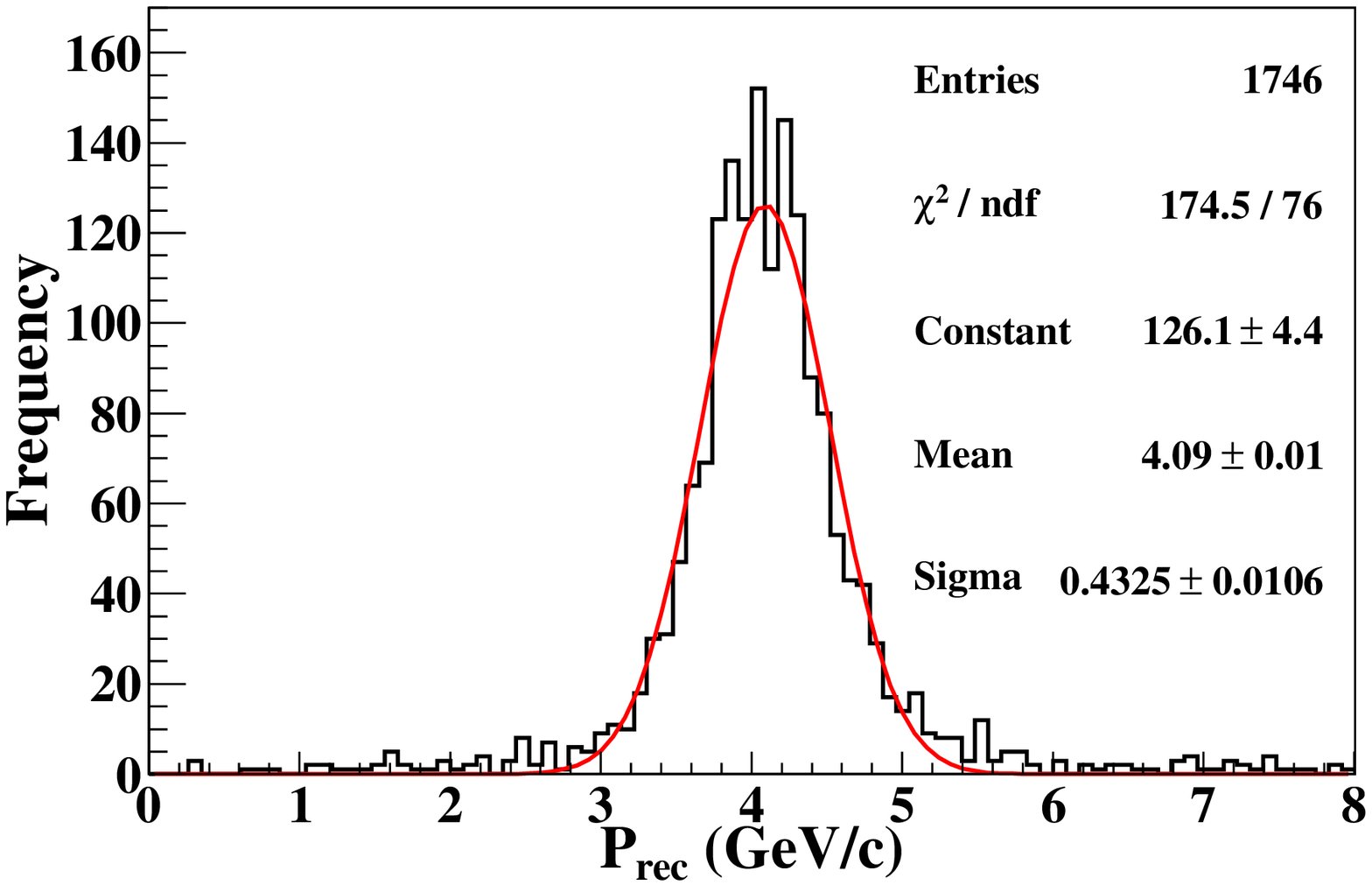}}
  \subfloat[$ 3\pi/4 < \mid \phi \mid \le \pi
$]{\label{fig:55d}\includegraphics[totalheight=.18\textheight,
width=0.50\textwidth]{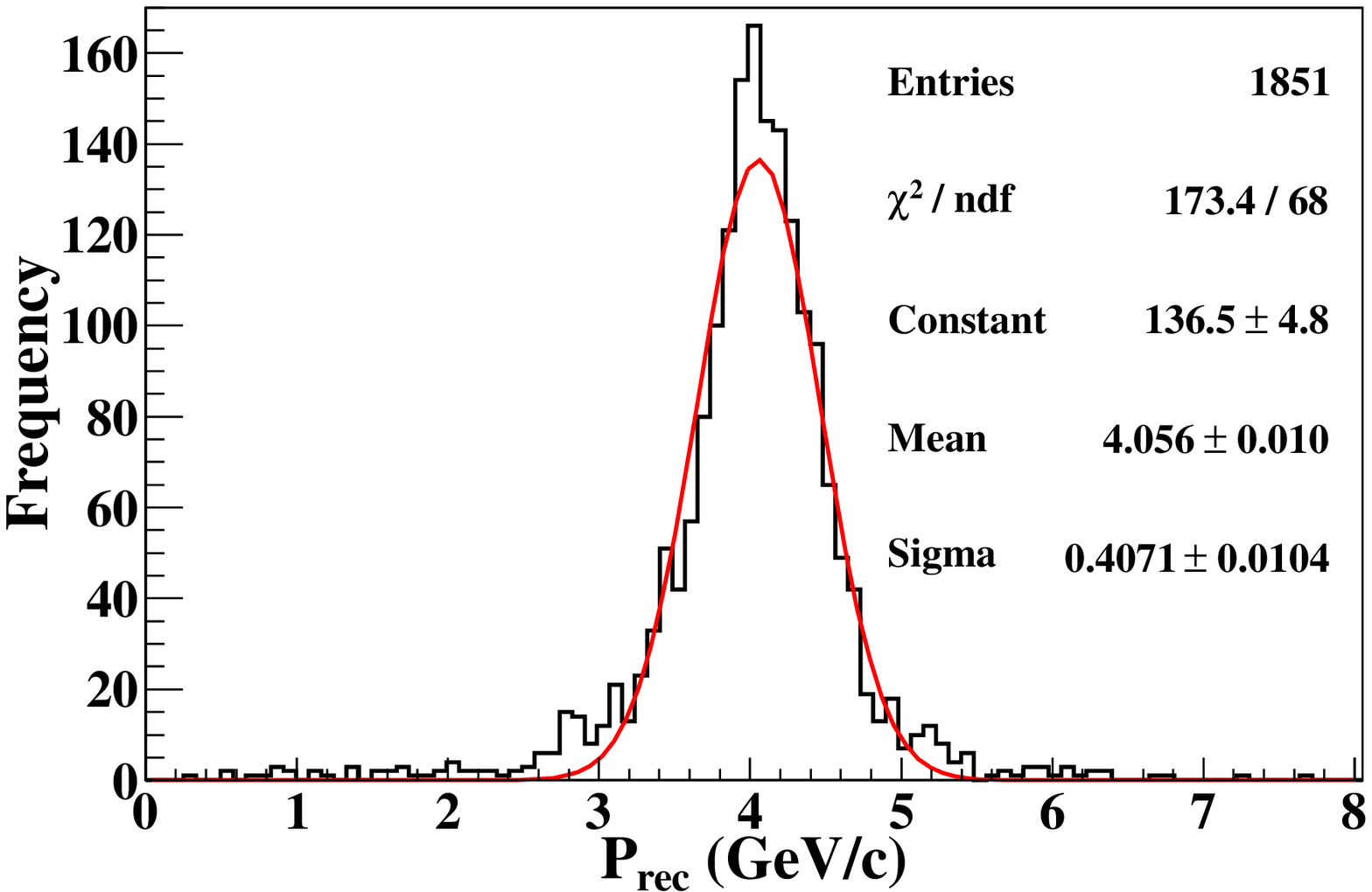}}
\caption{Reconstructed momentum distribution for $(P_{\rm in}, \cos\theta) =
(4\hbox{ GeV/c}, 0.65)$ in different $\phi$ bins, fitted with a
Gaussian distribution.} 
\label{fig:sv_4_65}
\end{figure}

{{Fig.~\ref{fig:sv_4_65} shows the Gaussian fits for the
reconstructed histogram of the reconstructed momentum $P_{\rm rec}$ for
input values of $P_{\rm in}=4$ GeV/c and $\cos\theta=0.65$ in different
$\phi$ regions.}}

{{As can be seen, the quality of fits are reasonable. Both
the mean and the width of the distributions for the four sets vary;
the former by about $4\sigma$ across all sets, and the latter by more than
$10\sigma$. This is why the fits to the combined data sets are relatively
poor ($\chi^2/{\hbox{ndf}}=8$) while, on separating these events, the fits
are vastly improved, with good fits and acceptable $\chi^2/{\hbox{ndf}}$
in each region. In particular, marginally more events are reconstructed
in phi bins I and IV due to the effect of the magnetic field, while the
resolution is best in bins III and IV, as discussed earlier.}}

\begin{figure}[btp]
\renewcommand{\figurename}{Fig.}
  \centering
  \subfloat[cos$\theta$ =
0.45]{\label{fig:12a}\includegraphics[totalheight=0.18\textheight,width=0.45\textwidth]{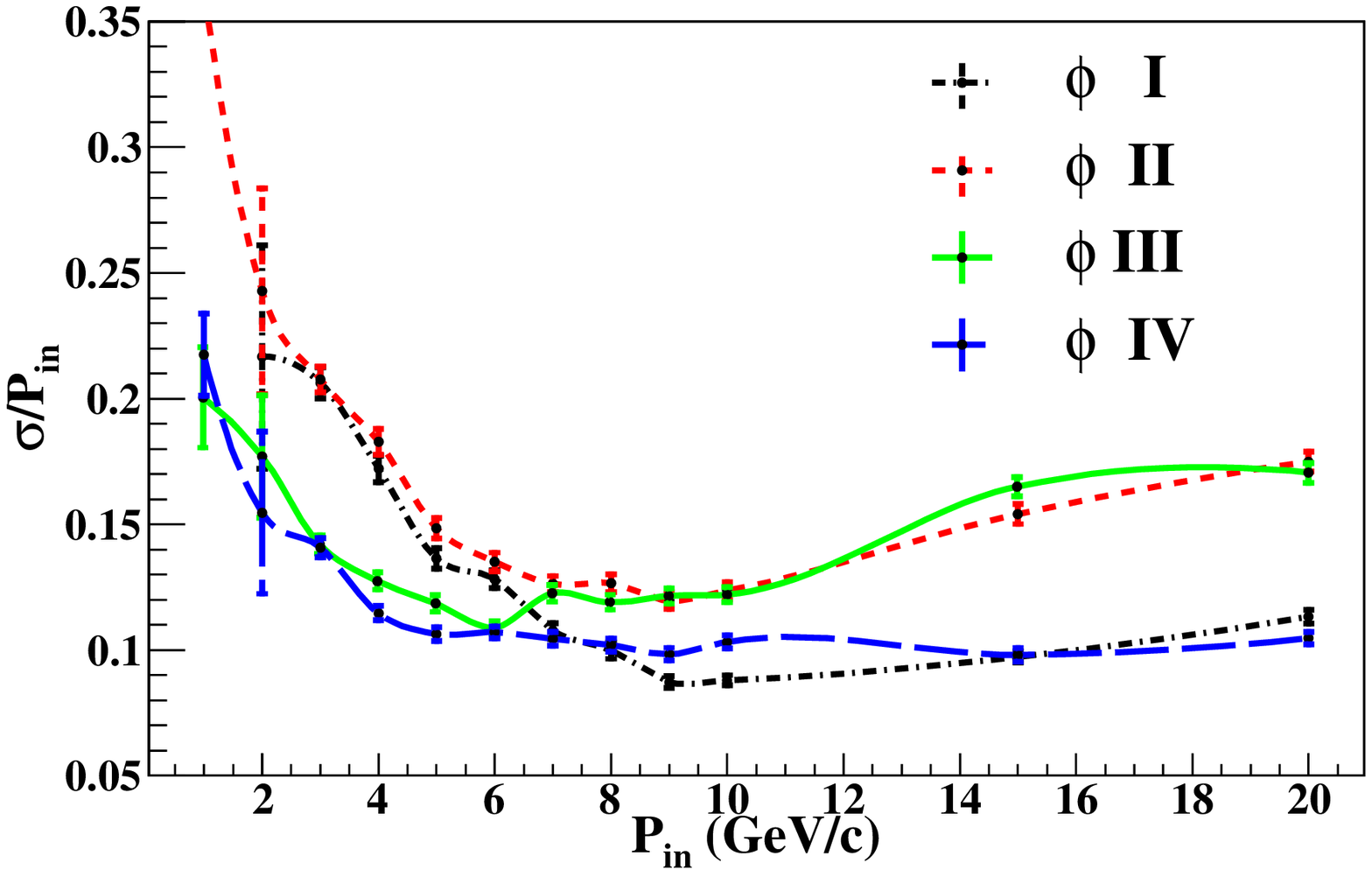}}
  \subfloat[cos$\theta$ =
0.65]{\label{fig:12b}\includegraphics[totalheight=0.18\textheight,width=0.45\textwidth]{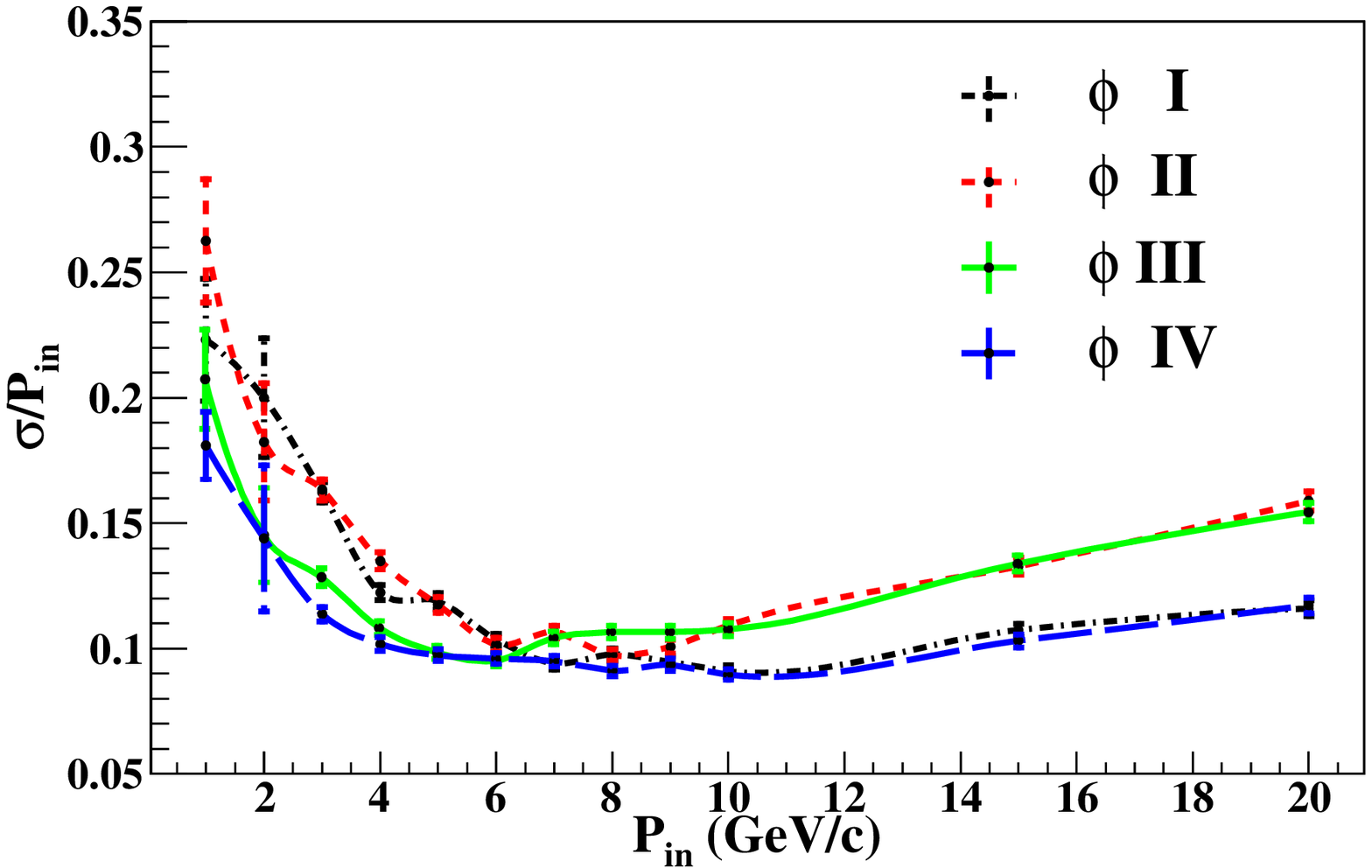}}\\
  \subfloat[cos$\theta$ =
0.85]{\label{fig:12c}\includegraphics[totalheight=0.18\textheight,width=0.45\textwidth]{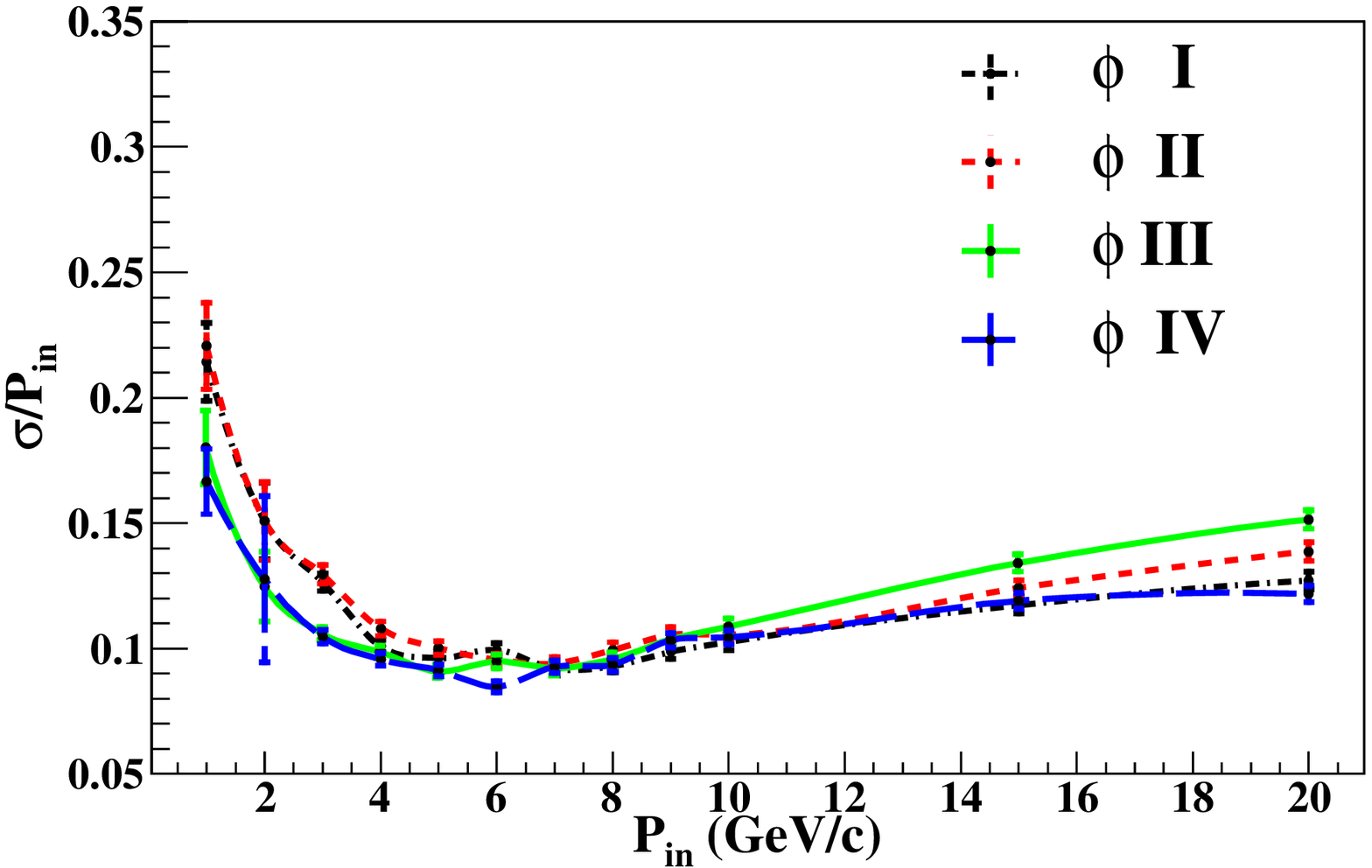}}
  \caption{Muon resolution as a function of muon momentum for different
(fixed) values of $\cos\theta=0.45, 0.65, 0.85$,
shown for different $\phi$ bins.}
\label{fig:sv_E_ct}
\end{figure}

The momentum resolution as a function of input momentum for different
values of $\cos\theta$ in the different $\phi$ bins is shown in
Fig.~\ref{fig:sv_E_ct} for momenta from 2--20 GeV/c. It can be seen
that initially (upto about 6 GeV/c), the resolution improves with
increasing energy. This is because, at such small momenta, as the
momentum increases, the number of RPC layers crossed by the muons also
increases, thus increasing the number of layers and hits; in addition,
the magnetic field causes appreciable bending so the accuracy of momentum
reconstruction also increases. Thus the resolution improves.

As the input momentum increases further, the particle begins to exit the
detector, so that only a partial track is contained in the detector;
this comprises relatively straight sections since the radius of curvature
increases with momentum, resulting in a poorer fit. Due to this the
resolution then worsens as the input energy increases. In particular,
muons in bins II and III exit the detector from the $y$-direction
earlier than those in bins I and IV due to the larger $x$ dimension
of the detector. Hence, at higher energies, the resolutions in bins
I and IV are better than those in bins II and III.

{{In contrast, at low energies, where the tracks are completely contained,
muons in bins III and IV with $\vert\phi \vert> \pi/2$ have negative
$x$-component of momentum, and, as explained
earlier, have better resolution than muons with the same $(P_{\rm in},
\cos\theta)$ but with $\vert\phi\vert < \pi/2$ as in bins I and II. Note
however that roughly equal number of events are reconstructed in each
$\phi$ bin.}}

At low momenta, $P_{\rm in} < 5$ GeV/c, when the track is completely
contained, it is possible to determine the momentum purely from range
measurements since the muon is a minimum ionising particle. While the
results (for the resolution $R$) are somewhat better than the fits
obtained from the Kalman filter technique, they are very unreliable
at small angles because of the high probability of losses in detector
dead spaces. Hence, the use of range to determine the track momentum
is limited and can be applied on a case-by-case basis, depending on the
location of the track. However, we do not discuss this further here.

\begin{figure}[btp]
\renewcommand{\figurename}{Fig.}
  \centering
  \subfloat[$ \vert \phi \vert \le \pi/4
$]{\label{fig:11a}\includegraphics[totalheight=.20\textheight,
width=0.50\textwidth]{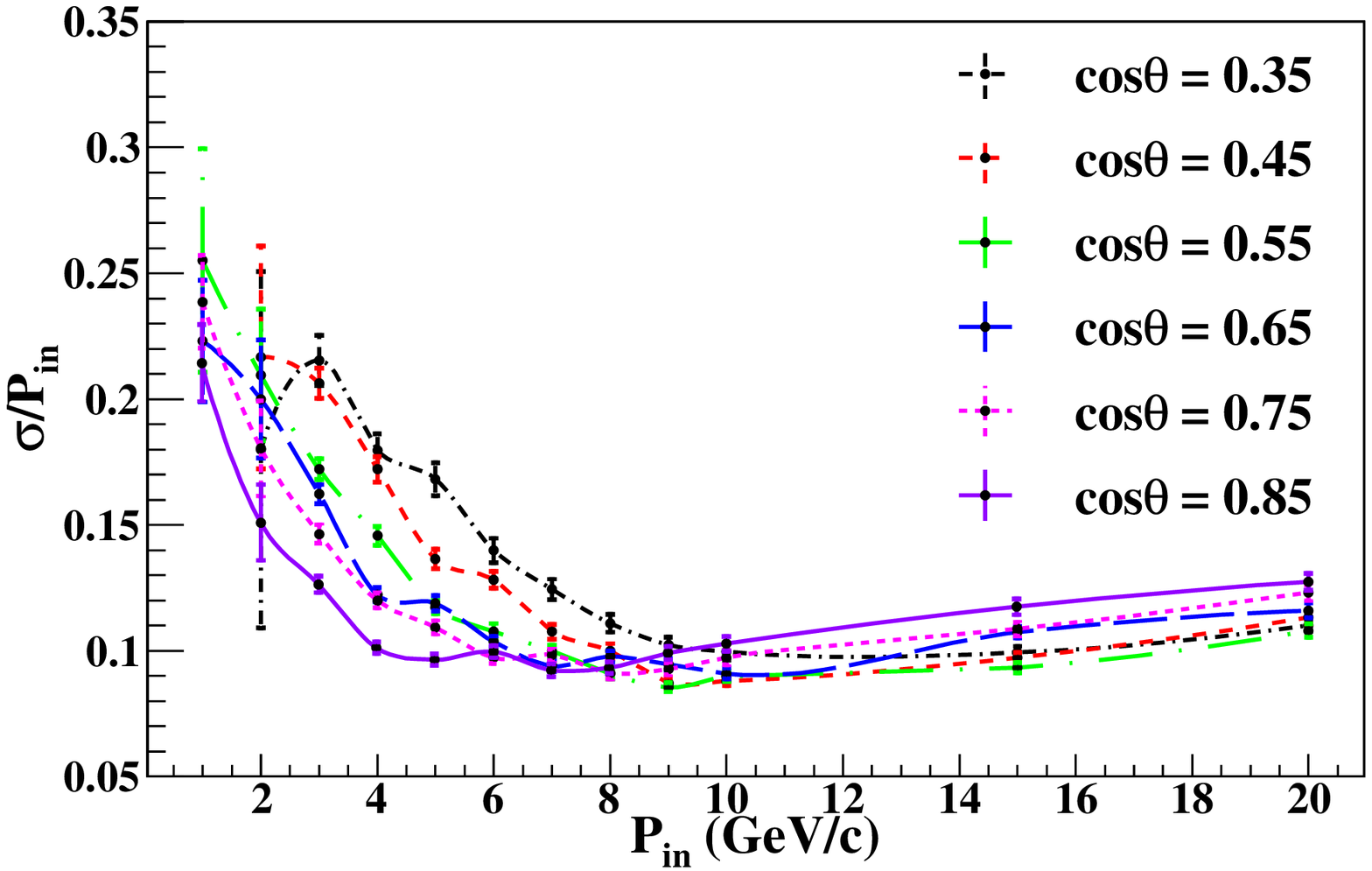}}
  \subfloat[$ \pi/4 < \vert \phi \vert \le \pi/2
$]{\label{fig:11b}\includegraphics[totalheight=.20\textheight,
width=0.50\textwidth]{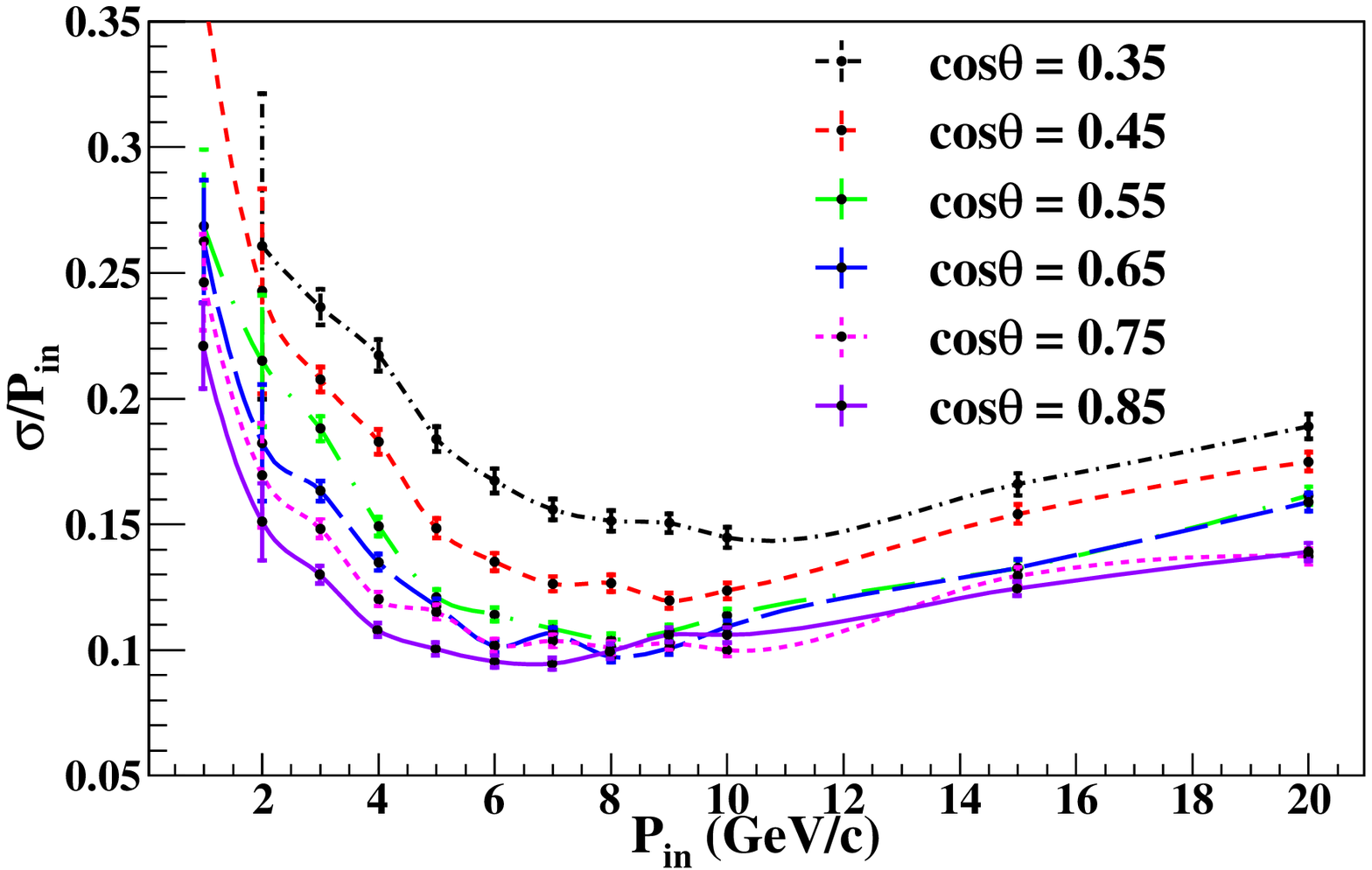}}\\
  \subfloat[$ \pi/2 < \vert \phi \vert \le 3\pi/4
$]{\label{fig:11c}\includegraphics[totalheight=.20\textheight,
width=0.50\textwidth]{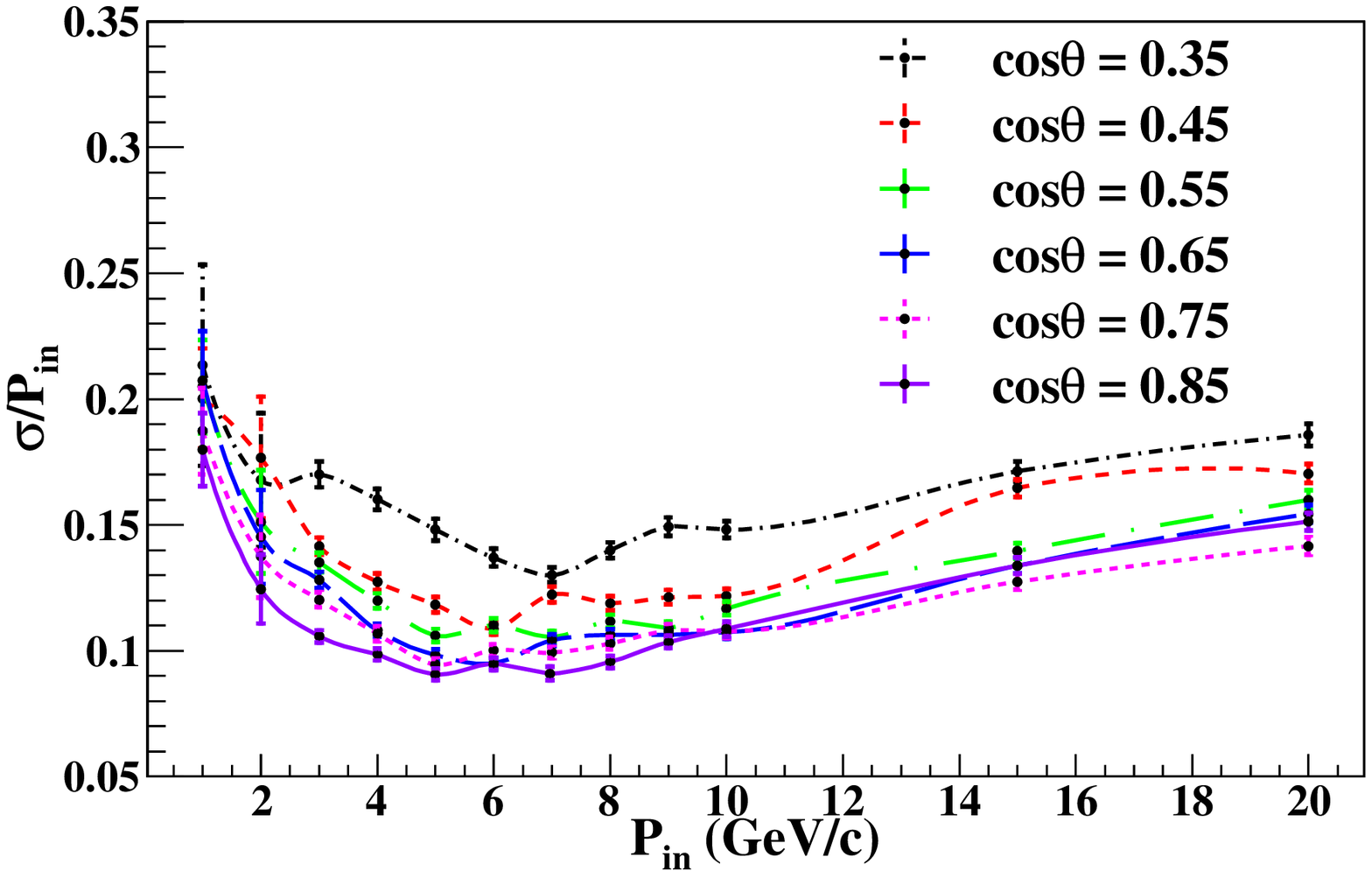}}
  \subfloat[$ 3\pi/4 < \vert \phi \vert \le \pi
$]{\label{fig:11d}\includegraphics[totalheight=.20\textheight,
width=0.50\textwidth]{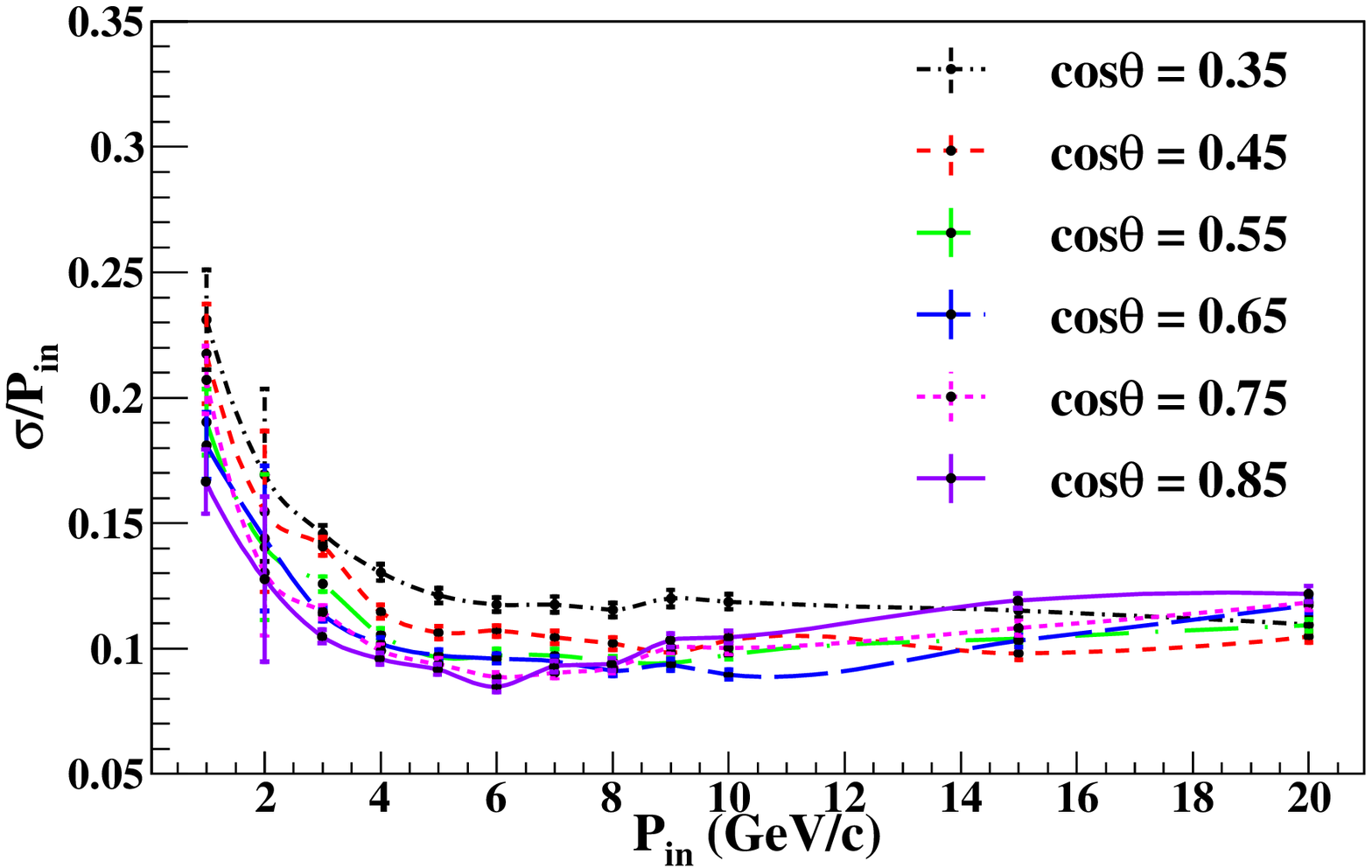}}
\caption{Muon resolution as a function of input momentum and
$\cos\theta$, in different bins of azimuthal angle $\phi$.}
\label{fig:sv_E_ct_phi}
\end{figure}

Finally, the
variation of resolution with $P_{\rm in}$ for a range of $\cos\theta$
are shown in Fig.~\ref{fig:sv_E_ct_phi}.

In general, at angles larger than about $70^\circ$ ($\cos\theta = 0.35$),
the resolution is relatively poor since there are several times fewer
hits than at more vertical angles. This is a consequence of the detector
geometry, with its horizontal layers of iron plates where energy is
dominantly lost. Hence ICAL is not so sensitive to very horizontal muons.

In the next section, we describe the other parameters of interest such
as the mean shift, reconstruction effciency, etc. These are not very
sensitive to the azimuthal angle; hence, in what follows we combine all
the $\phi$ bins and calculate $\phi$-averaged quantities.

\subsection{Mean Shift}

\begin{figure}[htp]
\centering
\renewcommand{\figurename}{Fig.}
\includegraphics[width=0.80\textwidth]{{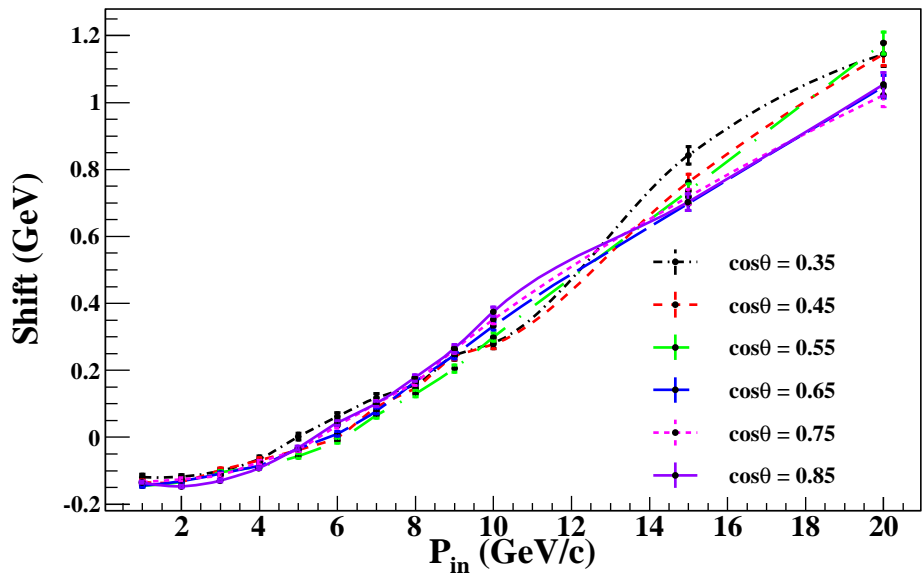}}
\caption{Shift in the mean of reconstructed momentum as a function of the input momentum.} 
\label{fig:Meanshift}
\end{figure}

The mean of the reconstructed momentum ($P_{\rm rec}$) distribution is approximately independent of both the polar and azimuthal angles.
Fig.~\ref{fig:Meanshift} shows the shift in the mean of reconstructed
momentum $(\hbox{shift} = P_{\rm in} - P_{\rm rec})$, as a function of the
input momentum. The shift can arise due to multiple scattering, etc.,
and is roughly linear beyond a few GeV/c.

\noindent 

\subsection{Momentum Reconstruction Efficiency}

\begin{figure}[htp]
\renewcommand{\figurename}{Fig.}
\begin{center}
\includegraphics[width=0.8\textwidth, height=0.5\textwidth]{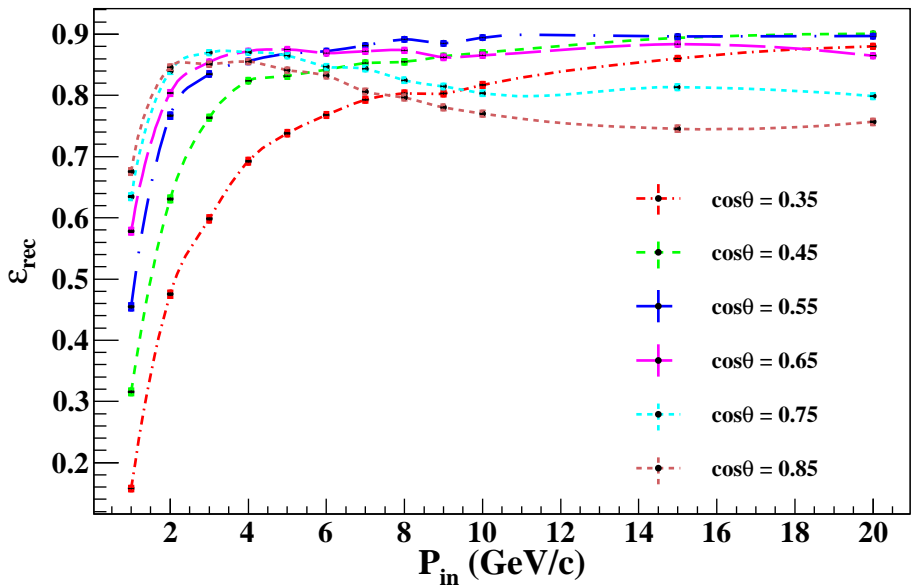}
\end{center}
\caption{Momentum reconstruction efficiency as a function of the input momentum
for different $\cos\theta$ values.}
\label{fig:2}
\end{figure}

The momentum reconstruction efficiency ($\epsilon_{\rm rec}$) is defined as the
ratio of the number of reconstructed events, $n_{\rm rec}$ (irrespective
of charge), to the total number of generated events, $N_{total}$. We have
\begin{eqnarray}
\epsilon_{\rm rec} & = & \frac{n_{\rm rec}} {N_{\rm total}}~, \\ \nonumber
{\hbox{and its error, }} \delta \epsilon_{\rm rec} & = & \sqrt{\epsilon_{\rm rec}(1-\epsilon_{\rm
rec})/N_{\rm total}}~.
\end{eqnarray} 
Fig.~\ref{fig:2} shows the muon momentum reconstruction efficiency as a function
of input momentum for different $\cos\theta$ bins. The momentum
reconstruction efficiency is somewhat smaller than the track
reconstruction efficiency, especially at smaller angles. It is seen that the
efficiency of momentum reconstruction depends on the energy of
the incident particle, the strength of the magnetic field, the angle of
propagation, etc.

For momentum values below 4 GeV/c as the input momentum
increases, the reconstruction efficiency increases for all angles,
since the number of hits increases as the particle crosses more number
of layers. At larger angles, the reconstruction efficiency for small
energies is smaller compared to vertical angles since the number of hits
for reconstructing tracks is less. But as the input energy increases,
since the particle crosses more number of layers, the efficiency of
reconstructing momentum also increases and becomes comparable with
vertical angles. At higher energies the reconstruction efficiency becomes
almost constant. The drop in efficiency at high energies for vertical
muons is due to the track being partially contained as well as the
requirement of single track being reconstructed. Note that a similar
figure in Ref.~\cite{physics} does not include this requirement, and
therefore corresponds to track reconstruction efficiency as defined in
this paper.

\subsection{Relative Charge Identification Efficiency}

\begin{figure}[htp]
\renewcommand{\figurename}{Fig.}
\begin{center}
\includegraphics[width=0.8\textwidth, height=0.5\textwidth]{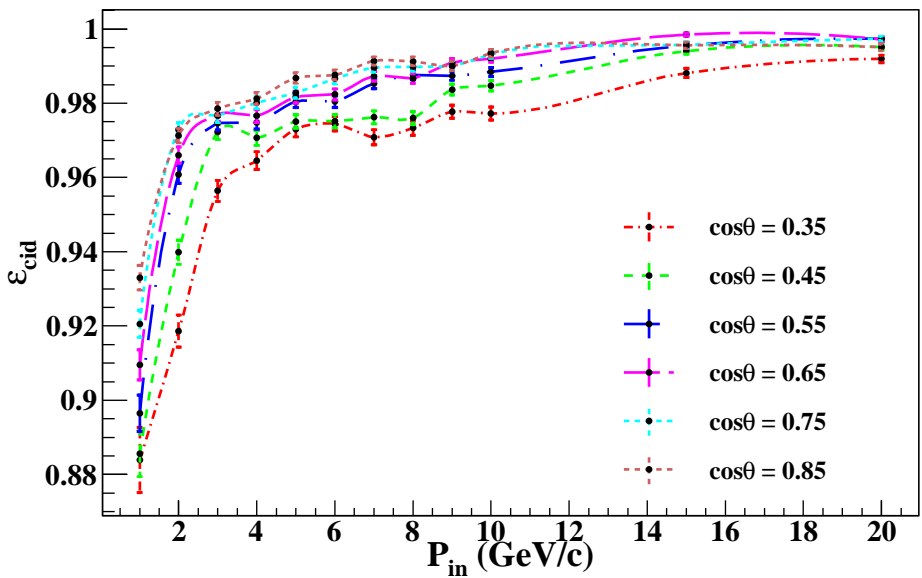}
\end{center}
\caption{The relative charge identification efficiency as a function of
the input momentum for different $\cos\theta$ values. Note that the $y$-axis
range is 0.87--1 and does not start from zero.}
\label{fig:4}
\end{figure}

The charge identification of the particle plays a crucial role in the
determination of the neutrino mass hierarchy since it distinguishes
neutrino from anti-neutrino induced events. The charge of the
particle is determined from the direction of curvature of the track in
the magnetic field. Relative charge identification efficiency is defined
as the ratio of number of events with correct charge identification,
$n_{\rm cid}$, to the total number of reconstructed events, i.e.,
\begin{eqnarray}
\epsilon_{\rm cid} & = & \frac{n_{\rm cid}} {n_{\rm rec}}~,\\ \nonumber
\hbox{with error, } \delta \epsilon_{\rm cid} & = &
    \sqrt{\epsilon_{\rm cid}(1-\epsilon_{\rm cid})/n_{\rm rec}}~.
\end{eqnarray}
Note that the total reconstructed events also includes those which are
reconstructed in the wrong $\theta$ quadrant. Fig.~\ref{fig:4} shows
the relative charge identification efficiency as a function of input
momentum for different polar angles $\cos\theta$. Here we consider the
number of reconstructed events after applying the additional selection
criterion. A marginally worse charge identification efficiency is obtained
for vertical events ($\cos\theta = 0.85$) when this additional criterion
is not applied, while results at other angles are virtually the same.

The muon undergoes multiple scattering during propagation in the detector;
for small momentum, since the number of layers with hits is small, this
may lead to an incorrectly reconstructed direction of bending, resulting
in the wrong charge identification. Hence the charge identification
efficiency is relatively poor at lower energies, as can be seen from
Fig.~\ref{fig:4}. As the energy increases the length of the track
also increases due to which the charge identification efficiency also
improves. Beyond a few GeV/c, the charge identification efficiency becomes
roughly constant, about 98--99\%. While this is expected to hold up to
$P_{\rm in} \sim 50$ GeV/c or more, the momentum resolution at such high
momenta will be much worse; but this is outside the scope of the present
study.

\section{Discussions and Conclusion}

The ICAL detector is proposed as the primary detector to be installed in
the INO cavern, and its main goal is the study of neutrino oscillation
parameters through atmospheric neutrinos. The detector is mainly
sensitive to charged-current events from interactions of atmospheric
neutrinos with detector material (mostly iron) where muons are produced
(sometimes with additional hadrons as well), depending on the type and
nature of the interaction. Hence it is crucial to correctly understand
the response of ICAL to muons by detailed simulations and this is the focus
of this work. The ICAL geometry was simulated using GEANT4 software
and muons with fixed momenta from 1--20
GeV/c and and with direction $\cos\theta > 0.35$ 
were propagated through this simulated detector and their
characteristics studied. In the current study, muons were generated in
what is called the central part of the ICAL detector where the
magnetic field is large and uniform. Both contained and partially
contained (with muons exiting the detector) events were analysed.

It was found that clean tracks were obtained in the detector for muons
with momenta from few to 10s of GeV/c
in all zenith angles smaller than $\cos\theta = 0.35$. The
momentum of the muons was reconstructed with the help of a Kalman filter
algorithm that makes use of the bending of the tracks in the magnetic
field. Hence the presence of the magnetic field is crucial to this
study. 

Different $\phi$ bins were analyzed separately in order to find the
dependence of resolution on azimuthal angle. This was motivated by the
presence of the magnetic field as well as the coil gaps and support
structures that break
the azimuthal symmetry, although neutrino oscillations as well as the
neutrino flux (at the relevant energies considered here) are independent
of $\phi$, a strong east-west effect ($\phi=0$ versus $\phi=\pi$) being
visible \cite{honda} only at sub-GeV energies.

Each sample was analysed for the momentum resolution, reconstruction
efficiency, charge identification efficiency and direction
resolution. While the momentum resolution was about 9--14\%, the
momentum
reconstruction efficiency was better than about 80\% in most of the regions. The direction resolution was
found to be indeed very good, being better than a degree for all
angles for momenta greater than about 4 GeV/c, which is most important
for studying the neutrino mass hierarchy through matter effects. The
relative charge identification efficiency was also about 98\% over
this range.

In summary, ICAL simulations indicate that the detector has a good
response to muons, including identifying their momentum, direction,
and charge, with good efficiency. These results have already been used to
perform the physics analysis of atmospheric neutrinos with ICAL
\cite{physics}, \cite{physics1} and are now being used in many
other physics analyses.

\section*{Acknowledgments}
We thank the INO collaboration members and simulations group including
P. Behera, Md. Naimuddin for criticism and comments; G. Majumder and
A. Redij for code implementation, code-related discussions and many
clarifications; K. Bhattacharya and N. Dash for discussions; B.S. Acharya,
S. Chattopadhyay for a critical reading of the manuscript and useful
comments, and S. Behera for discussions on the magnetic field map. KR
acknowledges DST/UGC (Govt. of India) for financial support.

\section*{Appendix A: The Kalman Filter Algorithm}

In the Kalman filter algorithm a track is represented by a set of
parameters called state vector which contains the information about
the position of the hits, direction of the track and the momentum
of the particle at that position. A state vector is defined as $x =
(x, y, \d x/\d z, \d y/\d z, q/p)$. The state covariance matrix, $C$,
a $5 \times 5$ matrix, contains the expected error in the state vector.
The state vector is updated in every next plane using the information
about the current state vector. The estimated state vector is given by
$$
x_k  = F_{k-1} x_{k-1}+u_{k-1}~,
$$
where $F$ is the \textit{propagator matrix} which transports the state
vector from $(k-1)^{th}$ plane to $k^{th}$ plane and \textit{u} is the
process noise which contains information about multiple scattering and
energy loss by the particle. The propagator matrix contains the information
about the magnetic field and it is calculated for every pair of detector
planes. Track fitting is done to get the accurate estimate of the state
vector at each plane.

The propagator matrix, which has been simplified for the fast calculation
during track fitting \cite{Marshall}, depends on the difference in $z$
coordinates between two layers, the magnetic field values and the
information about the energy loss of the particle. The form of the
propagator matrix is
$$
F=
  \begin{bmatrix}
    1 & 0 & \delta z & 0 & \frac{1}{2} B_y \delta z^2\\
    0 & 1 & 0 & \delta z & -\frac{1}{2} B_x \delta z^2\\
    0 & 0 & 1 & 0 & B_y \delta z\\
    0 & 0 & 0 & 1 & -B_x \delta z\\
    0 & 0 & 0 & 0 & 1 + \epsilon\\
  \end{bmatrix}
$$
If the hits are in successive layers, $\delta z= 9.6$ cm,
which is the separation between adjacent RPCs in the $z$-direction. The
magnetic field values at each position are calculated by interpolating
the field map while $\epsilon$ takes into account the energy loss by
the particle as given by the Bethe formula \cite{Bethe}.

The noise covariance matrix takes into account the uncertainities due to
multiple scattering and energy loss by the particle which is in the form
$$
Q=
  \begin{bmatrix}
    \textbf{Q}^{\textbf{MS} } & \textbf{0}\\
    0 & Q^{\delta E}\\
  \end{bmatrix}
$$
where $\textbf{Q}^{\textbf{MS} }$ is a $4 \times 4$ matrix which
contains the information about the uncertainities in multiple scattering
as given in Ref.~\cite{Wolin} and $Q^{\delta E}$, a single term,
represents the uncertainites in the energy loss. It is assumed that the
process noise is unbiased.

The actual measurement at the \textit{$k^{th}$} plane, $m_k$, is a
function of the state vector:
$$
m_k  = H_k x_k+w_k~,
$$
where $H_k$ is the projection matrix and $w_k$ is the measurement noise
(with covariance matrix $V_k$). Here
$m_k$ is a column matrix which contains the (digitised) strip
information from $x$ and $y$ pick-up panels for each plane i.e., ($x_k,
y_k$); the strip width of $1.96$ cm gives an error of about $\pm 1$
cm in each of these two directions.

The Kalman prediction equation for the $k^{th}$ layer is,
$$
x_{k}^{k-1}  = F_{k-1} x_{k-1}^{k-1}~,
$$
and so the best estimate of the state vector including the $k^{th}$
measurement is given as
$$
x_{k}^{k}  = x_{k}^{k-1}+K_k(m_k-H_k x_{k}^{k-1})~,
$$
where $K_k$, the Kalman gain matrix, is used to decide the relative
importance to give to the prediction and to the measurement. The filter
iteratively shrinks the error in the estimates by processing the input
hits, one after another. The process converges very well once at least 5
hits are passed to the fitter program.

\section*{Appendix B: Effect of Dead Spaces}

The dead spaces, namely the coil gap and the support structures that
have no active detector elements, are more or less uniformly distributed
throughout ICAL. In order to understand the impact of these on the
detector resolution, we consider a sample where the muon vertices are
essentially fixed at the
location $(x,y,z) = (100 \hbox{ cm},100 \hbox{ cm},0 \hbox{ cm})$ which
is equidistant from the supports at $x, y = 0, \pm 200$ cm (with a very
small smearing over a small volume $(\pm 10 \hbox{ cm}, \pm 10\hbox{ cm}, \pm
10\hbox{ cm})$).

\begin{figure}[htp]
\renewcommand{\figurename}{Fig.}
  \centering
\includegraphics[width=0.49\textwidth,height=0.3\textwidth]{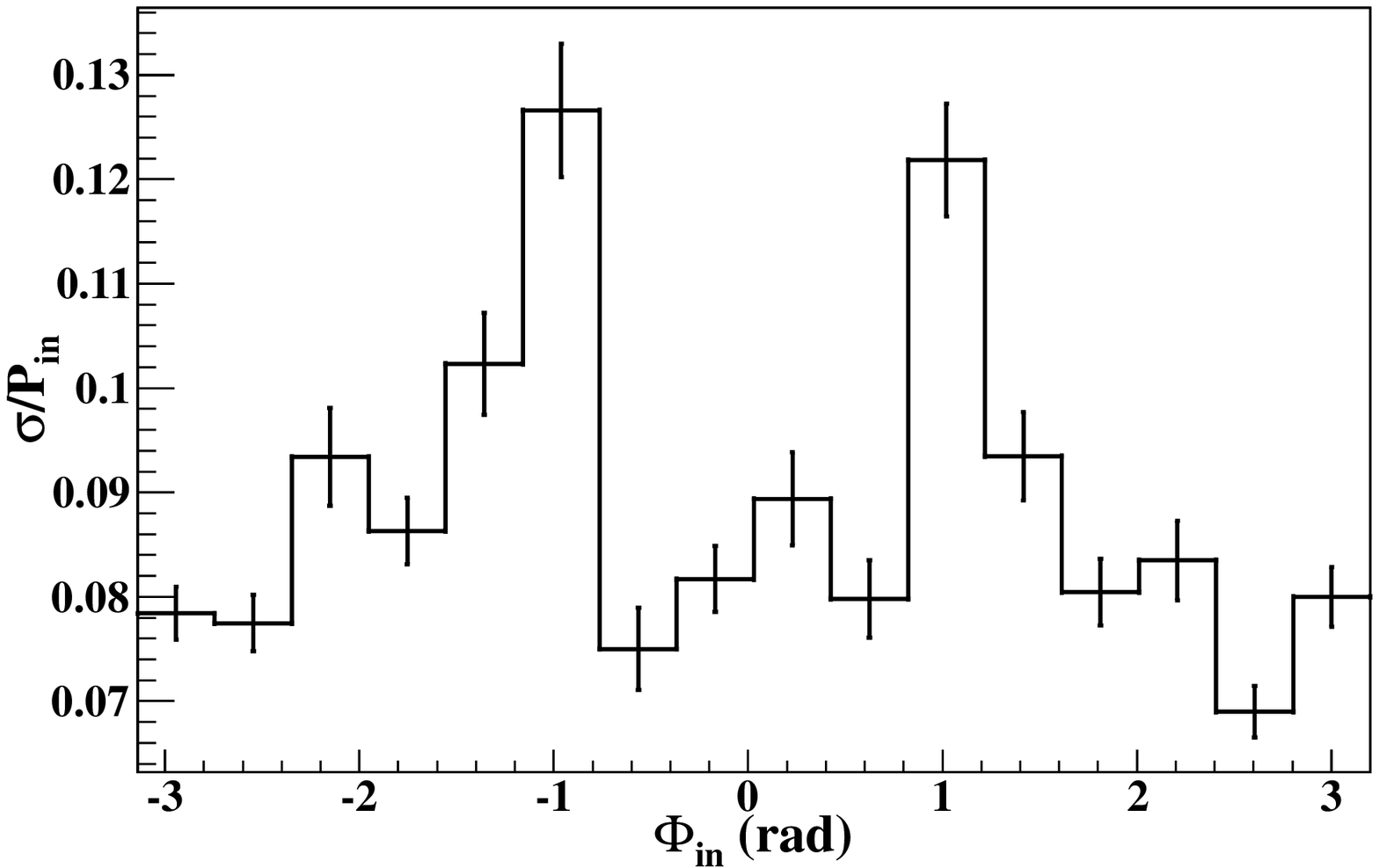}
\includegraphics[width=0.49\textwidth,height=0.3\textwidth]{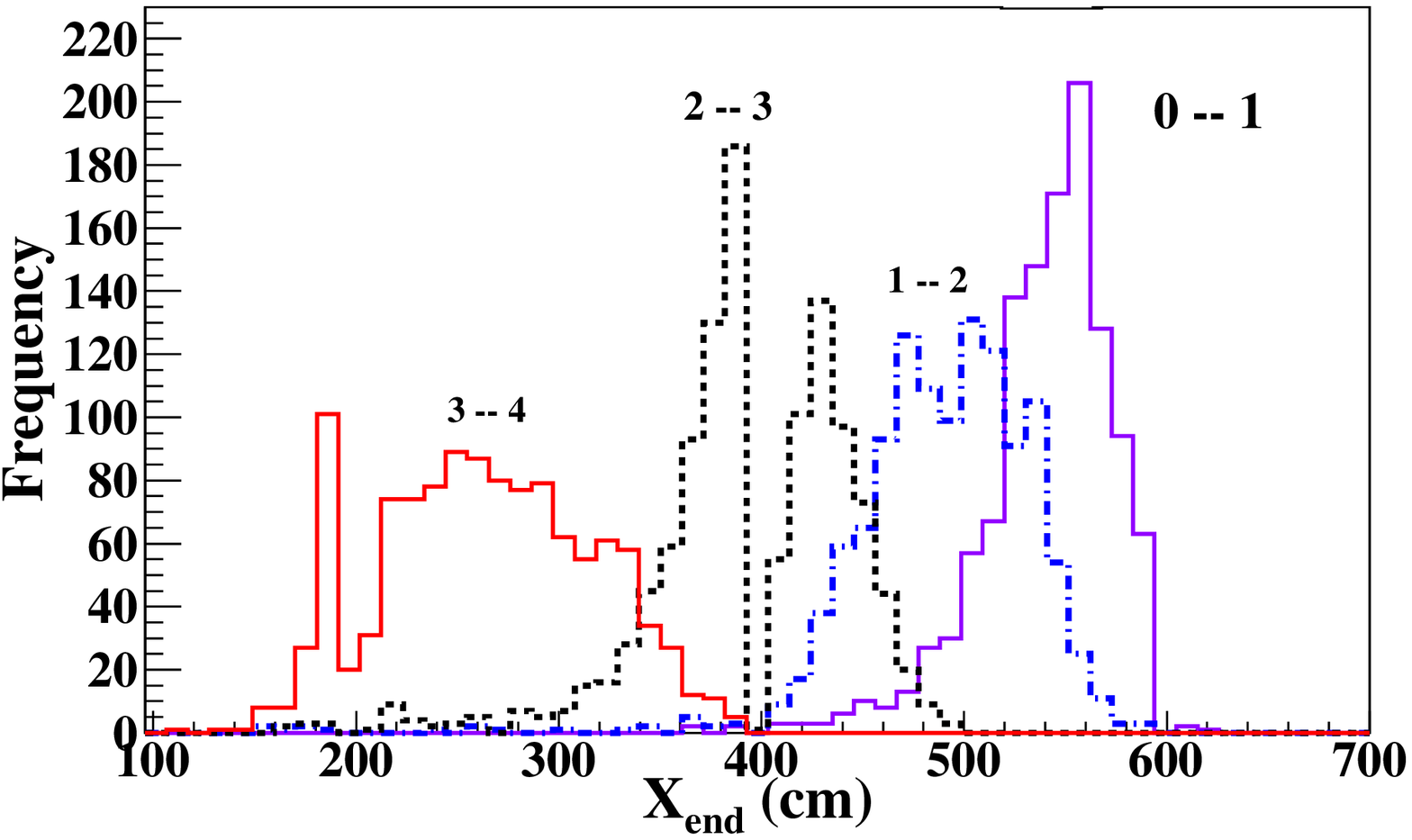}
\caption{Left panel: Relative momentum resolution, $\sigma/P_{in}$,
for muons with fixed momentum $(P_{in}, \cos\theta) = (5\hbox{ GeV/c},
0.65)$, plotted in bins of the azimuthal angle $\phi$, with $\phi=0$
corresponding to the $x$-direction.
Right panel: Distributions of end $x$ positions of the track
in four sample bins of azimuthal angle.}
\label{fig:phidep}
\end{figure}

Fig.~\ref{fig:phidep} shows the momentum resolution $R = \sigma/P_{\rm
in}$, as a function of the azimuthal angle $\phi$. The response is
symmetric in $\phi \longleftrightarrow -\phi$, within errors. Furthermore,
the best response corresponds to the events where the muon's in-plane
momentum is along the $x$ axis, i.e., $\vert \phi \vert \sim 0, \pi$,
while, the worst resolution is for muons in the $\phi$ bin 2--3 ($\times
\pi/8$). This can be understood as follows. A survey of the end-point
positions of the tracks show that the worst resolution corresponds to the
bin where most of the tracks end in the coil gaps. This can be seen in
the right panel of Fig.~\ref{fig:phidep} where end-$x$ values of muons,
$x_{\rm end}$, with input $\vert \phi \vert$ values in the bins (0--1,
1--2, 2--3, and 3--4) $\times \pi/8$ are shown (that for other values of
$\phi$ are suppressed for clarity); it is seen that a large fraction of
muons in the $\phi$ bin 2--3 $\times \pi/8$ end in the coil gap ($x=400$
cm). In addition, the resolutions worsen somewhat when more number of
tracks in the bin end in the support structures $x, y = 0, \pm 200,
\pm 400, \cdots$ cm, compared to other $\phi$ bins.

\begin{thebibliography}{99}

\bibitem{Athar:2006yb} M.S. Athar et al., 2006 India-based Neutrino Observatory: Project Report Volume I,
http://www.ino.tifr.res.in/ino/OpenReports/INOReport.pdf.

\bibitem{physics} T.~Thakore, A.~Ghosh, S.~Choubey and A.~Dighe, JHEP {\bf 1305}, 058 (2013). 

\bibitem{physics1}A.~Ghosh, T.~Thakore and S.~Choubey, JHEP {\bf 1304}, 009 (2013). 

\bibitem{asmita} INO Collaboration, {\it ICAL simulation using GEANT tool-kit}, in preparation, 2014.

\bibitem{geant} S. Agostinelli et al. {\it Geant4 - a simulation toolkit},
Nucl. Instr. \& Meth. in Phys. Res. {\bf A506} (2003) 250-303;
http://geant4.cern.ch/.

\bibitem{magnetcode} Infolytica Corp., Electromagnetic field simulation software, http://www.infolytica.com/en/products/magnet/.

\bibitem{rpc_char} B. Satyanarayana, {\it Design and Characterisation
Studies of Resistive Plate Chambers}, PhD thesis, (Department of Physics,
IIT Bombay, PHY-PHD-10-701, 2009).

\bibitem{hadronresponse} M.M. Devi et al., {\it Hadron energy response
of the ICAL detector at INO}, JINST {\bf 8} (2013) 11003.

\bibitem{kalman} R.E. Kalman, {\it A new approach to linear filtering and
prediction problems}, Journal of Basic Engineering {\bf 82} (1) (1960) 35--45.

\bibitem{Marshall} J.S. Marshall, {\it A study of muon neutrino
disappearance with the MINOS detectors and the NuMI neutrino beam},
PhD. Thesis, Univ. of Cambridge (2008).

\bibitem{Wolin} E. Wolin and L. Ho, {\it Covariance Matrices for Track Fitting
with the Kalman Filter}, Nucl. Instrum. Methods {\bf A329} (1993) 493.

\bibitem{Bethe} J. Beringer et al. (Particle Data Group), Phys. Rev.
{\bf D86} (2012) 010001; http://pdg.lbl.gov.

\bibitem{peripheral} INO Collaboration, {\it Muon response in the
peripheral region}, in preparation, 2014.

\bibitem{honda} Morihiro Honda, Takaaki Kajita, Katsuaki Kasahara, Shoichi
Midorikawa, Phys. Rev. {\bf D83} (2011) 123001.

\end{thebibliography}
\end{document}